\documentclass[aps,prx,reprint,showpacs,floatfix,amsmath,amssymb,superscriptaddress,longbibliography,footinbib]{revtex4-2}

\usepackage{graphicx}
\usepackage{multirow}
\usepackage{amsmath}
\usepackage{comment}
\usepackage{physics}
\usepackage{bm}
\usepackage{here}
\usepackage[colorlinks=true,urlcolor=blue,citecolor=blue, linkcolor=blue]{hyperref}
\begin{document}

\newcommand{\Yb}[1]{$^{#1}\mathrm{Yb}$}
\newcommand{\state}[3]{$^{#1}\mathrm{#2}_{#3}$}

\newcommand{\Ra}{\Rightarrow}
\newcommand{\ra}{\rightarrow}
\newcommand{\lra}{\longrightarrow}

\newcommand{\too}{\longrightarrow}
\newcommand{\mapstoo}{\longmapsto}

\newcommand{\df}{\mathrel{:=}}
\newcommand{\fd}{\mathrel{=:}}


\newcommand{\Prob}{\mathop{\textrm{Prob}} }
\newcommand{\Det}{\mathop{\textrm{Det}} }
\newcommand{\Pf}{\mathop{\textrm{Pf}} }
\newcommand{\sgn}{\mathop{\textrm{sgn}} }
\newcommand{\diag}{\mathop{\textrm{diag}} }
\renewcommand{\Re}{\mathop{\textrm{Re}} }
\renewcommand{\Im}{\mathop{\textrm{Im}} }

\newcommand{\Group}[2]{{ \hbox{{\itshape{#1}}($#2$)} }}
\newcommand{\U}[1]{\Group{U\kern0.05em}{#1}}
\newcommand{\SU}[1]{\Group{SU\kern0.1em}{#1}}
\newcommand{\SL}[1]{\Group{SL\kern0.05em}{#1}}
\newcommand{\Sp}[1]{\Group{Sp\kern0.05em}{#1}}
\newcommand{\SO}[1]{\Group{SO\kern0.1em}{#1}}

\newcommand{\scr}[1]{ \ensuremath{\mathcal{#1}} }

\newcommand{\sub}[1]{^{\vphantom{\dagger}}_{#1} }
\newcommand{\rsub}[1]{ \mathstrut_{\hbox{\scriptsize #1}} }
\newcommand{\rsup}[1]{ \mathstrut^{\hbox{\scriptsize #1}} }

\newcommand{\twi}{\widetilde}
\newcommand{\mybar}[1]%
    {{\kern 0.8pt\overline{\kern -0.8pt#1\kern -0.8pt}\kern 0.8pt}}
\newcommand{\roughly}[1]%
    {{ \mathrel{\raise.3ex\hbox{ $#1$\kern-.75em\lower1ex\hbox{$\sim$}} } }}
\newcommand{\Bra}[1]{\left\langle #1 \left|}
\newcommand{\Ket}[1]{\right| #1 \right\rangle}
\newcommand{\Tvev}[1]{\langle 0 | T #1 | 0 \rangle}
\newcommand{\avg}[1]{\langle #1 \rangle}
\newcommand{\Avg}[1]{\left\langle #1 \right\rangle}
\newcommand{\nop}[1]{:\kern-.3em#1\kern-.3em:}

\providecommand{\abs}[1]{\lvert#1\rvert}
\providecommand{\norm}[1]{\lVert#1\rVert}

\newcommand{\vacbra}{{\bra 0}}
\newcommand{\vac}{{\ket 0}}
\newcommand{\lsim}{\mathrel{\roughly<}}
\newcommand{\gsim}{\mathrel{\roughly>}}
\newcommand{\myint}{\int\mkern-5mu}
\newcommand{\ddx}[2]{d^{#1}#2\,}
\newcommand{\ddp}[2]{\frac{d^{#1}#2}{(2\pi)^{#1}}\,}
\newcommand{\del}{\partial}
\newcommand{\ddel}[2]{ \frac{\partial{#1}}{\partial{#2}} }
\newcommand{\delfb}{\overleftrightarrow{\partial}}
\newcommand{\al}{\ensuremath{\alpha}}
\newcommand{\be}{\ensuremath{\beta}}
\newcommand{\ga}{\ensuremath{\gamma}}
\newcommand{\Ga}{\ensuremath{\Gamma}}
\newcommand{\de}{\ensuremath{\delta}}
\newcommand{\De}{\ensuremath{\Delta}}
\newcommand{\ep}{\ensuremath{\epsilon}}
\newcommand{\ze}{\ensuremath{\zeta}}
\newcommand{\et}{\ensuremath{\eta}}
\renewcommand{\th}{\ensuremath{\theta}}
\newcommand{\Th}{\ensuremath{\Theta}}
\newcommand{\vth}{\ensuremath{\vartheta}}
\newcommand{\ka}{\ensuremath{\kappa}}
\newcommand{\la}{\ensuremath{\lambda}}
\newcommand{\La}{\ensuremath{\Lambda}}
\newcommand{\rh}{\ensuremath{\rho}}
\newcommand{\si}{\ensuremath{\sigma}}
\newcommand{\Si}{\ensuremath{\Sigma}}
\newcommand{\ta}{\ensuremath{\tau}}
\newcommand{\Up}{\ensuremath{\Upsilon}}
\newcommand{\ph}{\ensuremath{\phi}}
\newcommand{\vph}{\ensuremath{\varphi}}
\newcommand{\Ph}{\ensuremath{\Phi}}
\newcommand{\ch}{\ensuremath{\chi}}
\newcommand{\ps}{\ensuremath{\psi}}
\newcommand{\Ps}{\ensuremath{\Psi}}
\newcommand{\om}{\ensuremath{\omega}}
\newcommand{\Om}{\ensuremath{\Omega}}
\newcommand{ \eV}{ \ensuremath{\mathrm{ ~eV}} }
\newcommand{\keV}{ \ensuremath{\mathrm{~keV}} }
\newcommand{\MeV}{ \ensuremath{\mathrm{~MeV}} }
\newcommand{\GeV}{ \ensuremath{\mathrm{~GeV}} }
\newcommand{\TeV}{ \ensuremath{\mathrm{~TeV}} }
\newcommand{ \cm}{ \ensuremath{\mathrm{ ~cm}} }
\newcommand{\degree}{^{\circ}}
\newcommand{\dof}{DOF}
\newcommand{\nlsm}{\ensuremath{\mathrm{NL}\Sigma \mathrm{M}} }
\newcommand{\hc}{\mathrm{H.c.}} 
\newcommand{\n}{\notag \\}
\newcommand{\mcl}[1]{\mathcal{#1}}
\newcommand{\mfk}[1]{\mathfrak{#1}}
\newcommand{\mbb}[1]{\mathbb{#1}}
\newcommand{\fb}[1]{\overleftrightarrow{#1}}
\newcommand{\Z}{\phantom{-1}}
\newcommand{\obar}[1]{\overline{#1}}
\newcommand{\pfrac}[2]{\left(\frac{#1}{#2}\right)}

\title{Observation of nonlinearity of generalized King plot in the search for new boson}
\author{Koki Ono}
\altaffiliation{Electronic address: koukiono3@yagura.scphys.kyoto-u.ac.jp}
\affiliation{Department of Physics, Graduate School of Science, Kyoto University, Kyoto 606-8502, Japan}
\author{Yugo Saito}
\affiliation{Department of Physics, Graduate School of Science, Kyoto University, Kyoto 606-8502, Japan}
\author{Taiki Ishiyama}
\affiliation{Department of Physics, Graduate School of Science, Kyoto University, Kyoto 606-8502, Japan}
\author{Toshiya Higomoto}
\affiliation{Department of Physics, Graduate School of Science, Kyoto University, Kyoto 606-8502, Japan}
\author{Tetsushi Takano}
\affiliation{NICHIA Corp., 3-13-19, Moriya-Cho, Kanagawa-Ku, Yokohama, Kanagawa 221-0022, Japan}
\author{Yosuke Takasu}
\affiliation{Department of Physics, Graduate School of Science, Kyoto University, Kyoto 606-8502, Japan}
\author{Yasuhiro Yamamoto}
\affiliation{National Centre for Nuclear Research, Pasteura 7, Warsaw 02-093, Poland}
\affiliation{Physics Division, National Center for Theoretical Sciences, Taipei
10617, Taiwan}
\author{Minoru Tanaka}
\affiliation{Department of Physics, Graduate School of Science, Osaka University, Osaka 560-0043, Japan}
\author{Yoshiro Takahashi}
\affiliation{Department of Physics, Graduate School of Science, Kyoto University, Kyoto 606-8502, Japan}
\date{\today}
\begin{abstract}
We measure isotope shifts for neutral Yb isotopes on an ultranarrow optical clock transition \state{1}{S}{0}--\state{3}{P}{0} with an accuracy of a few Hz. 
The part-per-billion precise measurement was possible by loading the ultracold atoms into a three-dimensional magic-wavelength optical lattice and alternately interrogating the isotope pairs, thus minimizing the effects due to the optical lattice light-shift and inter-atomic interaction as well as the drifts of a clock laser frequency and a magnetic field.
The determined isotope shifts, combined with one of the recently reported isotope-shift measurements of Yb$^+$ on two optical transitions, allow us to construct the
King plots. 
Extremely large nonlinearity with the corresponding $\ch^2$ on the order of $10^4$ is revealed, and is not explained by a quadratic field shift. 
We further carry out the generalized King plot for three optical transitions so that we can eliminate the contribution arising from a higher-order effect within the Standard Model 
which might explain the observed nonlinearity of King plots for two transitions. 
Our analysis of the generalized King plot shows a deviation from linearity at the 3$\sigma$ level, indicating that there exist at least two higher order contributions in the measured isotope shifts. Then, under the reasonable assumption to attribute them to higher-order field shifts within the Standard Model, we obtain the upper bound of the product of the couplings for a new boson mediating a force between electrons, and neutrons $|y_ey_n|/(\hbar c)< 1\times10^{-10}$ for the mass less than 1~keV with the 95\% confidence level is derived, providing an important step towards probing new physics via isotope-shift spectroscopy.

\end{abstract}
\maketitle

\section{\label{Intro}Introduction}

The standard model (SM) of particle physics offers an excellent explanation of most of the phenomena in nature~\cite{Tanabashi2018}.
It is, however, an empirical model, therefore it does not provide any explanation about the three generations of matter and the origin of mass mixing. 
In addition, some phenomena have escaped from the proper explanation by the SM such as cosmological phenomena~\cite{Bertone2005, Dine2003} including dark matter and matter-antimatter asymmetry, as well as the strong CP problem~\cite{Kim2010}. 
There have been continuous efforts in the search for physics beyond the SM in various experiments ranging from high-energy frontier~\cite{Tanabashi2018} to low-energy precision measurements~\cite{Safronova2018}.

 

 
Recently, a novel proposal of detecting a new boson beyond the SM mediating a new force between neutrons and electrons attracts considerable attention~\cite{Delaunay2017, Berengut2018}. 
Such a coupling between neutrons and electrons will manifest itself in an isotope-dependent resonant frequency shift of electronic transitions, which will be detected spectroscopically. 
However, due to the large uncertainty in the atomic energy calculation which requires scarcely known nuclear properties such as a charge distribution within a nucleus, it is hopeless to directly compare the experimentally determined resonance frequency with the theoretical prediction for each isotope. 
To overcome this difficulty, the proposal relies on the linear relation of the isotope shifts (ISs) of two different electronic transitions, known as King plot linearity~\cite{King1963, King2013}, which should hold under the usually accepted assumption that the IS consists of mass-shift (MS) and field-shift (FS) terms each of which is expressed as a product of nuclear-dependent and electronic-transition-dependent factors. 
Introducing the new extra isotope-dependent term, called the particle shift (PS), results in a nonlinearity of the King plot.
The PS can be described by the Yukawa potential $V(r) =(-1)^{s+1} y_e y_n \exp(-m c r/\hbar)/(4\pi r)$, where $m$ and $s$ stand for the mass and the spin of the new boson, respectively, and $y_e (y_n)$ are the couplings of the new boson to electron(neutron). Here $c$ and $\hbar$ represent the speed of light and the reduced Planck constant, respectively.

Motivated by this proposal, so far, the precise IS measurements for Ca$^{+}$~\cite{Knollmann2019, Solaro2020}, Yb$^{+}$~\cite{Counts2020}, Sr$^{+}$~\cite{Manovitz2019}, and Sr~\cite{Takano2017, Miyake2019} are recently reported.
In particular, impressive precision of about 10~mHz in the IS measurement is demonstrated for a particular isotope pair of $^{87}$Sr--$^{88}$Sr by a state-of-the-art optical lattice clock technique~\cite{Takano2017}, and for a pair of $^{86}$Sr$^{+}$--$^{88}$Sr$^{+}$ using a novel two-isotope entangled state~\cite{Manovitz2019}.
Important progress has been reported for Ca$^{+}$~\cite{Solaro2020,Knollmann2019} and Yb$^{+}$~\cite{Counts2020} where systematic precision IS measurements using two different optical transitions are performed and used to carry out a King plot analysis. 
While the resulting King plot for the Ca$^{+}$ IS data with about 20~Hz accuracy is consistent with the linearity, an evidence of nonlinearity at the three standard deviation level is observed in the King plot for the Yb$^{+}$ 300 Hz~precision IS data.
Although a new particle gives rise to the nonlinearity of the King plot in principle, higher-order effects within the SM can also result in nonlinearities~\cite{Blundell1987}, and thus limits the sensitivity to new physics~\cite{Mikami2017,Flambaum2018,Tanaka2019}.
It is noted that the higher-order terms in the mass shift are much smaller for heavy elements such as Yb \cite{Palmer1987,Flambaum2018}.
More recently it is argued that the theoretical analysis within the SM could explain the result of nonlinearity observed in Yb$^{+}$ by considering a quadratic field shift (QFS)~\cite{Counts2020}, the next leading order Seltzer moment or an isotope-dependent nuclear deformation~\cite{Allehabi2021}.
It is noted that such an analysis requires very accurate atomic and nuclear theory calculations, the validity of which should be carefully checked.    




To overcome the difficulty associated with this SM contribution, the generalization of the King plot is proposed in Ref.~\cite{Mikami2017}.
In this generalized King plot, the IS data for more than two electronic transitions are utilized to eliminate the SM contribution by considering the nonlinearity in dimensions higher than two. 
A numerical calculation was reported in a recent paper~\cite{Berengut2020}, with an ultranarrow (6s)$^{2}$~\state{1}{S}{0}--6s6p~\state{3}{P}{0} transition of a neutral Yb and already reported two transitions of Yb$^{+}$ as an example. 
Also the recent theory papers discuss the use of the Yb \state{1}{S}{0}--\state{3}{P}{0} clock transition~\cite{Dzuba2018,Schelfhout2021} as one of the transitions for the King plot.    
It is noted that absolute frequency measurement of the Yb \state{1}{S}{0}--\state{3}{P}{0} clock transition was reported with less than 1~Hz accuracy~\cite{Poli2008, Riehle2018} only for two isotopes of \Yb{171} and \Yb{174} and with 10~Hz accuracy for \Yb{173} \cite{Clivati2016}, and there has been no report on any measurements for five(or more) isotopes which is the minimum requirement in the construction of the generalized King plot.


Here we report the first systematic precise measurement of ISs for six neutral Yb isotopes including five bosons on an ultranarrow optical clock transition \state{1}{S}{0}--\state{3}{P}{0}.
By working with a large number of ultracold atoms loaded into a magic-wavelength optical lattice~\cite{Katori2003} where the polarizabilities in the \state{1}{S}{0} and \state{3}{P}{0} are quite close, we can largely suppress the light shift due to the optical lattice and the Doppler effect. 
In addition, the three-dimensional (3D) optical lattice configuration and the irradiation of photoassociation (PA) beam enable us to realize a system consisting of each atom isolated and localized in one lattice site with no multiple occupancy, thus free from a collisional shift. 
Furthermore, a measurement scheme of alternate interrogation of the isotope pairs minimizes the effects due to the drifts of a clock laser frequency and a magnetic field.
Various systematic effects such as a quadratic Zeeman shift, a residual optical lattice light shift, a probe light shift, and so on, are carefully examined.
As a result, the ISs are determined with an accuracy of a few Hz, corresponding to a more-than two orders of magnitude improvement over the recent Yb$^{+}$ measurement~\cite{Counts2020}.
It is noted that the accuracy of our measurement is checked by confirming the consistency between our IS measurement for the \Yb{171} and \Yb{174} pair and the absolute frequency measurements in NIST~\cite{Poli2008, Riehle2018} within the uncertainty of about 1~Hz.
The determined ISs are utilized to discuss the nonlinearity of King plots by combining the recently reported IS measurements of Yb$^{+}$ on \state{2}{S}{1/2}--\state{2}{D}{3/2} and \state{2}{S}{1/2}--\state{2}{D}{5/2} transitions.
The King plots using the \state{1}{S}{0}--\state{3}{P}{0} and one of the above-mentioned Yb$^{+}$ transitions show very large nonlinearity.
This demonstrates the advantage of using the two optical transitions associated with the electronic states of very different characters in obtaining a high sensitivity on the higher-order effect, owing to the lack of a cancellation mechanism.  
In order to eliminate the higher-order effect, we construct the generalized King plot by using the ISs for all three optical transitions.
Importantly, our analysis of the generalized King plot shows a deviation from linearity at the 3$\sigma$ level, and the upper bound of the product of the new boson couplings $|y_ey_n|/(\hbar c)<1\times10^{-10}$ for the new boson mass less than 1~keV with the 95\% confidence level (C.L.) is derived.
In addition, this work will also trigger theoretical efforts to discriminate between different nuclear models through theory-experiment comparisons.


\section{\label{IS}Isotope shift and linear relation}
\subsection{\label{Isotope shifts and 2D King plot}2D King plot}
The IS between the isotope pair of $(A',A)$ for the transition $\la$ can be parametrized in the good approximation as
\begin{align}
   \nu_{\la}^{A'A}
 =&
    K_{\la} \de\mu^{A'A} +F_{\la} \de\avg{r^2}^{A'A}.
\label{EqKing}
\end{align}
The two terms on the right-hand side are known as the leading order of the MS and FS.
Here, $\de\mu^{A'A}$ is the inverse mass difference of nuclei $1/m_{A'} -1/m_A$, $\de\avg{r^2}^{A'A}$ is the difference of the nuclear mean square charge radii $\avg{r^2}^{A'}-\avg{r^2}^A$ (see Appendix~\ref{mass and radius} for the values adopted in this work), and the isotope independent factors $K_{\la}$ and $F_{\la}$ are the electronic factors given by the electron density.


When we consider this leading order IS in Eq.~\eqref{EqKing} for two distinct transitions $\la_1$ and $\la_2$, we obtain a linear relation for $\nu_{\la_1}^{A'A}$ and $\nu_{\la_2}^{A'A}$, 
\begin{align}
  \nu_{\la_2}^{A'A} = F_{\la_2\la_1} \nu_{\la_1}^{A'A} +K_{\la_2\la_1} \de\mu^{A'A},
\label{Eq2d}
\end{align}
where $F_{\la_2\la_1}$ and $K_{\la_2\la_1}$ are the isotope independent coefficients, 
given as $F_{\la_2\la_1}=F_{\la_2}/F_{\la_1}$ and $K_{\la_2\la_1}=K_{\la_2} -F_{\la_2\la_1} K_{\la_1}$.
Here, ambiguous nuclear dependence $\de\avg{r^2}^{A'A}$ is eliminated between the transitions.
Equivalently, dividing Eq.~\eqref{Eq2d} by $\de\mu^{A'A}$, we obtain the original King linear relation for $\bar{\nu}_{\la_1}^{A'A}$ and $\bar{\nu}_{\la_2}^{A'A}$, 
\begin{align}
  \bar{\nu}_{\la_2}^{A'A} = F_{\la_2\la_1} \bar{\nu}_{\la_1}^{A'A} +K_{\la_2\la_1},
\label{Eq2dKing}
\end{align}
where $\bar{\nu}_{\la}^{A'A}\equiv\nu_{\la}^{A'A}/\de\mu^{A'A}$ is called the modified IS~\cite{King1963, King2013}.
In Eq.~\eqref{Eq2dKing}, the slope and intercept are isotope-independent, and thus all the ISs data are plotted along a single line in a $(\bar{\nu}_{\la_2}^{A'A}, \bar{\nu}_{\la_1}^{A'A})$ two dimensional (2D) space.

The higher-order IS term, denoted as $\nu_{\la,ho}^{A'A}$, can violate the linear relation~\cite{Blundell1987, Delaunay2017}.
For the higher-order IS within the SM, relevant for heavy elements like Yb~\cite{Blundell1987}, we consider two higher-order FSs,
\begin{align}
  \nu_{\la,ho}^{A'A} \rightarrow G_{\la}^{(4)} \de\avg{r^4}^{A'A} +G_{\la}^{(2)} [\de\avg{r^2}^2]^{A'A}, 
\label{EqNonlinear}
\end{align}
where the first term is the next leading order Seltzer moment, and the second one is the QFS.
The isotope dependence of the next leading order Seltzer moment is given by $\de\avg{r^4}^{A'A} =\avg{r^4}^{A'}-\avg{r^4}^A$~\cite{Seltzer1969}.
As introduced in Ref.~\cite{Counts2020}, that of the QFS can be described by $[\de\avg{r^2}^2]^{A'A} = (\de\avg{r^2}^{A'A_0})^2 -(\de\avg{r^2}^{A A_0})^2$, where $A_0$ is the reference nucleus.
This isotope dependence is simply obtained by the modification of the na\"{i}ve $(\de\avg{r^2}^{A'A})^2$ term to satisfy the transitive consistency condition of the IS $\nu_\la^{A'A} =\nu_\la^{A'A_0}+\nu_\la^{A_0 A}$.
They are obviously equivalent to the relations $[\de\avg{r^2}^2]^{A'A} =(\de\avg{r^2}^{A'A})^2$ if $A$ is chosen as the same nucleus with the reference $A_0 =A$.
Their electronic factors are written by $G_\la^{(4)}$ and $G_\la^{(2)}$, respectively.

 For the higher-order IS beyond the SM, we consider the PS, a possible new physics contribution given by the weakly interacting light boson, 
\begin{align}
 \nu_{\la,ho}^{A'A} \rightarrow  \al_\text{NP} X_{\la} (A' -A).
\label{EqNonlinear_PS}
\end{align}
For the PS, we assume the new physics contribution can be described by the Yukawa potential $V(r)$, given in Sec.~\ref{Intro}.
In Eq.~\eqref{EqNonlinear_PS}, we define the reduced coupling $\al_\text{NP}=(-1)^{s+1} y_e y_n/(4\pi\hbar c)$ and the electronic factor of the new physics as the expectation value of the potential
\begin{align}
X_\la=\frac{c}{2\pi} \int_0^\infty d r\de\rho_\la(r)\frac{e^{-\frac{m cr}{\hbar}}}{r},
\end{align}
where $\de\rho_\la(r)$ represents the change in the radial electron density function during the transition $\la$.

We need to include the higher-order ISs once the experimental precision reaches a level of the higher-order ISs.
In this case, $\nu_{\la}^{A'A}$ is given as 
\begin{align}
   \nu_{\la}^{A'A}
 =&
   K_{\la} \de\mu^{A'A} +F_{\la} \de\avg{r^2}^{A'A} +H_{\la} \de\et^{A'A},
\label{EqKingHO}
\end{align} 
which now includes either of the higher-order IS terms considered in the above as $H_{\la} \de\et^{A'A}$, where $\de\et^{A'A}$ and $H_{\la}$ are the isotope dependence and the electronic factor, respectively.
Then, the violation of the linear relation is formulated as,
\begin{align}
  \nu_{\la_2}^{A'A} = F_{\la_2\la_1} \nu_{\la_1}^{A'A} +K_{\la_2\la_1} \de\mu^{A'A} +H_{\la_2\la_1}\de\et^{A'A},
\label{Eq2d_NL}
\end{align}
where the last term is responsible for the nonlinearity, and $H_{\la_2\la_1}=H_{\la_2} -F_{\la_2\la_1} H_{\la_1}$.
If we employ the right source of the higher-order effect and its accurate isotope dependence, the data can be fit within errors.
We note that the isotope dependence of the higher-order effect is usually very difficult to evaluate.

\subsection{\label{Isotope shifts and 3D King plot}3D King plot}
This discussion of the King linearity is generalized to the case of more than two transitions, proposed in Ref.~\cite{Mikami2017}.
Here, we consider the case where we have the ISs of three different transitions of $\la_1$, $\la_2$, and $\la_3$, and four or more independent pairs of isotopes.
While the ISs are given as Eq.~~\eqref{EqKingHO}, the IS data for three transitions satisfy the following linear relation of
\begin{align}
\nu_{\la_3}^{A'A} = f_{\la_1} \nu_{\la_1}^{A'A} +f_{\la_2} \nu_{\la_2}^{A'A} +  k_\mu \de\mu^{A'A},
\label{Eq3d}
\end{align}
where $f_{\lambda_1}$, $f_{\lambda_2}$, and $k_\mu$ are the isotope independent coefficients associated with the three transitions.
They can be specified as in the case of Eq.~\eqref{Eq2d}. 
%
%
Dividing the Eq.~\eqref{Eq3d} by $\de\mu^{A'A}$, we now obtain the generalized King linear relation between $\bar{\nu}_{\la_1}^{A'A}$, $\bar{\nu}_{\la_2}^{A'A}$, and $\bar{\nu}_{\la_3}^{A'A}$ as 
\begin{align}
\bar{\nu}_{\la_3}^{A'A} = f_{\la_1} \bar{\nu}_{\la_1}^{A'A} +f_{\la_2} \bar{\nu}_{\la_2}^{A'A} +  k_\mu.
\label{Eq3dKing_mIS}
\end{align}
In this way, in addition to the leading-order FS term $\de\avg{r^2}^{A'A}$, a higher-order effect $\de\et^{A'A}$ is eliminated in the generalized 3D King plot while in the original 2D King linearity for two transitions only the $\de\avg{r^2}^{A'A}$ term is eliminated.
This strategy has a great advantage that we do not need to know the exact values for the isotope-independent electronic factors and isotope-dependent nuclear factors $\de\avg{r^2}^{A'A}$ and $\de\et^{A'A}$ which are often very difficult to evaluate.
The observation of nonlinearity of the generalized King plot Eq.~\eqref{Eq3dKing_mIS} has an important implication in the new particle search since the dominant higher-order effect which may well come from the origin within the SM should be eliminated. 
The observed nonlinearity can be analyzed using the relation 
\begin{align}
\nu_{\la_3}^{A'A} = f_{\la_1} \nu_{\la_1}^{A'A} +f_{\la_2} \nu_{\la_2}^{A'A} +  k_\mu \de\mu^{A'A} +h\de\zeta^{A'A},
\label{Eq3d_NL}
\end{align}
where the last term is responsible for the nonlinearity, originated from the further additional higher-order IS, and $h$ represents the isotope independent coefficient and $\de\zeta^{A'A}$ the nuclear dependent factor.
We give an explicit formula later. See Appendix~\ref{Formulation of h}.


\section{\label{Methods}Methods}

\begin{figure*}
\includegraphics[width=1\linewidth]{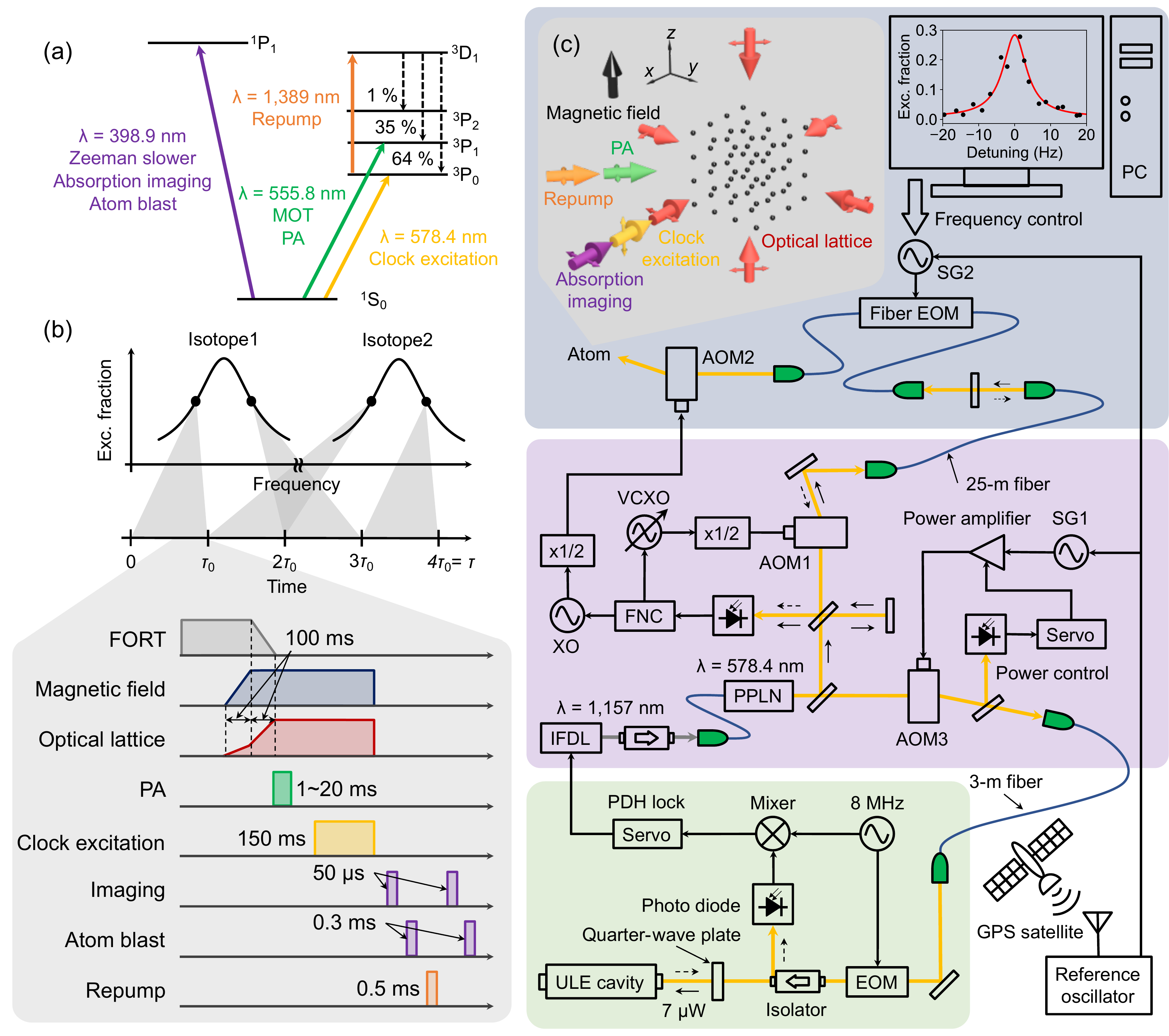}
\caption{Schematic diagram of experiment. (a) Relevant energy diagram of Yb atom. Note that the branching ratio for the radiative decay from \state{3}{D}{1} to the metastable state \state{3}{P}{2} is 1~\%~\cite{Porsev1999}, which is negligibly small compared with the measurement uncertainty, so we did not take this into account in our analysis. (b) The timing chart of measurement and pulse sequence after evaporative cooling. The pulse which removes remaining \Yb{174} atoms in the preparation of \Yb{176} or \Yb{171} samples is not included in this diagram. Note that a sequence time $\tau_0$ amounts to more than 100~s (see Appendix~\ref{Experimental parameters}), almost of which is spent for the loading of the atoms into a MOT from the Zeeman slower and the evaporative cooling. (c) Schematic illustration of experimental apparatus. An interference-filter-stabilized external-cavity diode laser (IFDL) at 1,156~nm is frequency doubled using a periodically poled lithium niobate (PPLN) waveguide to obtain the clock laser. The fiber-noise cancellation (FNC) is operated by controlling the voltage-controlled crystal oscillator (VCXO) driving the AOM1 so that the beatnote between the incident and returned light is phase-locked to the crystal oscillator (XO) driving AOM2. The signal generators SG1 and SG2 are referenced by the oven-controlled crystal oscillator disciplined by the Global Positioning System (GPS). The absorption images are analyzed by a PC, which controls the frequency of the SG2 driving a fiber EOM. The polarization of each lattice beam is linear and perpendicular to the magnetic field.}
\label{Schematics}
\end{figure*}

Our experiments start with the preparation of the ultracold atoms in a 3D optical lattice.
The basic sequence is the same for all five bosonic isotopes of \Yb{168}, \Yb{170}, \Yb{172}, \Yb{174}, and \Yb{176} used for the main IS measurements and two fermionic isotopes of \Yb{171} and \Yb{173} for the reference and investigation of the systematic effects. Figures \ref{Schematics}(a) and (b) show the Yb energy-level diagram relevant to the experiments and the pulse sequence, respectively. The atoms decelerated by the Zeeman slower with the \state{1}{S}{0}--\state{1}{P}{1} transition at 398.9~nm are captured by the magneto-optical trap (MOT) with the \state{1}{S}{0}--\state{3}{P}{1} transition at 555.8~nm. This is followed by loading the atoms into the crossed far-off-resonance trap (FORT), which is composed of the horizontal FORT at 1,064~nm and the vertical FORT at 1,070~nm. Subsequently, the atoms are cooled down to the nano-Kelvin regime by the evaporative cooling.
Since the $s$-wave scattering lengths of \Yb{171} and \Yb{176} atoms in the \state{1}{S}{0} state are not large enough for the thermalization with single species~\cite{Kitagawa2008}, they are cooled down using the sympathetic cooling with \Yb{174} atoms, which are later removed by the resonant light on the \state{1}{S}{0}--\state{3}{P}{1} transition after the evaporation.
The atoms are then loaded into the 3D optical lattice at the magic wavelength of $\lambda_\mathrm{m}=$759.349~nm~\cite{Barber2008}. The lattice depth for each axis is set to $30E_\mathrm{R}$, whose trap frequency is 22~kHz. Here $E_\mathrm{R}=k_\mathrm{B}\times95.4$~nK, with $k_\mathrm{B}$ being the Boltzmann constant, is the recoil energy for \Yb{174}. It is noted that the amplified spontaneous emissions from tapered amplifiers for the lattice beams are cut off by volume Bragg gratings with the resolution of less than 0.025~nm \cite{Fasano2021}. The lattice depth is calibrated with a pulsed lattice technique \cite{Denschlag2002}, and the lattice-laser beams pass through acousto-optic modulators (AOMs) for preventing interference between distinct lattice beams and for their power control and stabilization. The AOMs shift the lattice-laser frequencies for $x$, $y$ and $z$ axes by +80~MHz, -80~MHz and +85~MHz, respectively. A laser beam for PA, which is red detuned from the \state{1}{S}{0}--\state{3}{P}{1} resonance, is irradiated to remove the multiply occupied sites \cite{Akatsuka2008}. The isotope-dependent experimental parameters associated with the preparation of the cold atoms are summarized in Appendix~\ref{Experimental parameters}. 

After the PA, the interrogation pulse at 578.4~nm, with the Rabi frequency of 2$\pi\times$2.0~Hz, is applied for 150~ms in a magnetic field of $B_0=1.47$~mT in a typical condition. 
A finite magnetic field mixes the \state{3}{P}{1} state into the \state{3}{P}{0} state, allowing the \state{1}{S}{0}-\state{3}{P}{0} transition for the even isotopes, otherwise doubly forbidden \cite{Taichenachev2006}.
The clock excitation laser is stabilized using an ultra-low-expansion glass (ULE) cavity \cite{Takata2019} by means of the Pound-Drever-Hall (PDH) technique, as shown in Fig.~\ref{Schematics}(c), and the typical linewidth is a few Hz, confirmed by the beat measurement between two independently stabilized lasers. Since the clock excitation light is delivered to the optical table for the experiments via a 25-m optical fiber, the fiber-noise-cancellation (FNC) system by a phase-lock loop is configured \cite{Ma1994}.
Instead of adopting the offset locking to the ULE cavity for tuning the 578.4 nm clock laser frequency to each isotope resonance, the locking to the ULE cavity is performed at the fixed laser frequency to keep the steady locking operation, and
the frequency shift associated with the IS is introduced by a fiber electro-optic modulator (EOM) just before the atoms. The total intensity including the carrier and the sidebands amounts to $I_0=370$~mW/cm$^2$.

Finally, the numbers of the atoms excited to the \state{3}{P}{0} state as well as those remaining in the \state{1}{S}{0} are obtained by the consecutive absorption imaging so that we can extract the excitation fraction, rather than the excited atom number which is more sensitive to the atom number fluctuation. 
The first imaging pulse detects the \state{1}{S}{0} atoms, which is followed by removing the \state{1}{S}{0} atoms using the blast light. Then, the resonant light with the \state{3}{P}{0}--\state{3}{D}{1} transition at 1,389~nm is shed to repump the \state{3}{P}{0} atoms into the \state{1}{S}{0} state, and the second imaging pulse is applied \cite{Nemitz2016}. The \state{1}{S}{0}--\state{1}{P}{1} transition is employed for the imaging and the atom blast.

\begin{figure}
\includegraphics[width=0.8\linewidth]{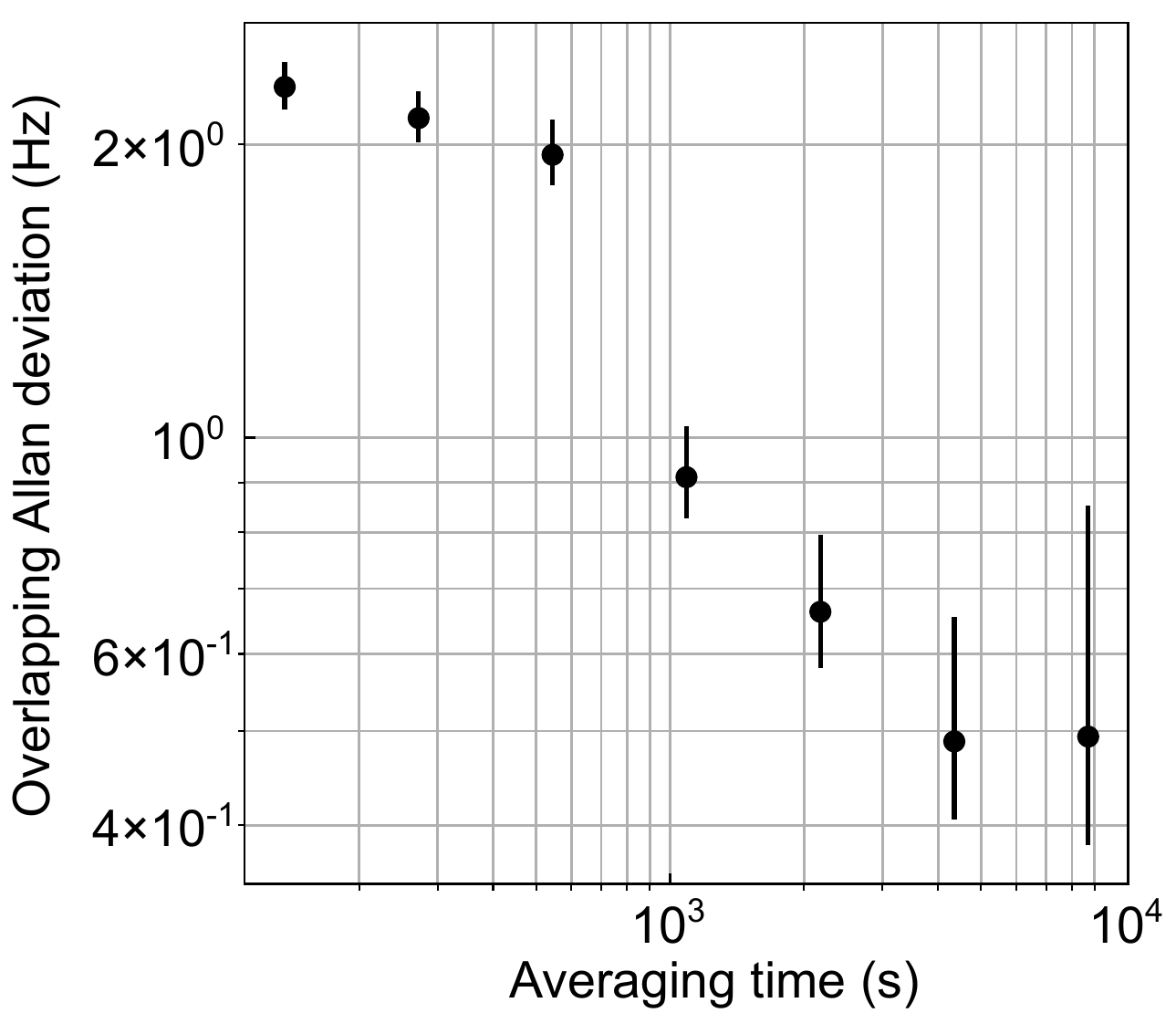}
\caption{Stability of interleaved clock operation. \Yb{173} atoms with $m_F=\pm5/2$ are alternately interrogated with a cycle time of $\tau=136$~s by the clock transition \state{1}{S}{0}($F=5/2,m_F=\pm5/2$)--\state{3}{P}{0}($F'=5/2,m_{F'}=\pm5/2$). Vertical and horizontal axes represent the overlapping Allan deviation and the averaging time, respectively. Error bars represent the upper and lower bounds with 1$\sigma$ confidence interval.}
\label{Adev}
\end{figure}

ISs are measured by the interleaved clock operation of the isotope pairs, which allows one to mitigate the systematic effects, such as the drifts of the clock-laser frequency and a magnetic field \cite{Takano2017}. Figure \ref{Schematics}(c) shows the schematic illustration of the clock operation, where two excitation fractions with different laser frequencies are compared and the obtained frequency-error signal is fed back to the signal generator driving the fiber EOM. The clock excitation of an isotope pair is performed alternately to implement the interleaved clock operation, and the IS is measured as the difference of the fiber EOM frequencies for the isotope pair. As shown in Fig.~\ref{Schematics}(b), the interrogations are repeated with a sequence time of $\tau_0$, and the cycle time of the IS measurement is $\tau = 4\tau_0$ since two measurements for each isotope are needed for a single IS measurement. Figure~\ref{Adev} shows the stability of the interleaved measurement, the total time of which amounts to 9~hours. The measured Allan deviation decreases with the increase of the averaging time, suggesting that the interleaved measurement does not suffer from serious long-term perturbations.

\begin{figure}
\includegraphics[width = 0.95\linewidth]{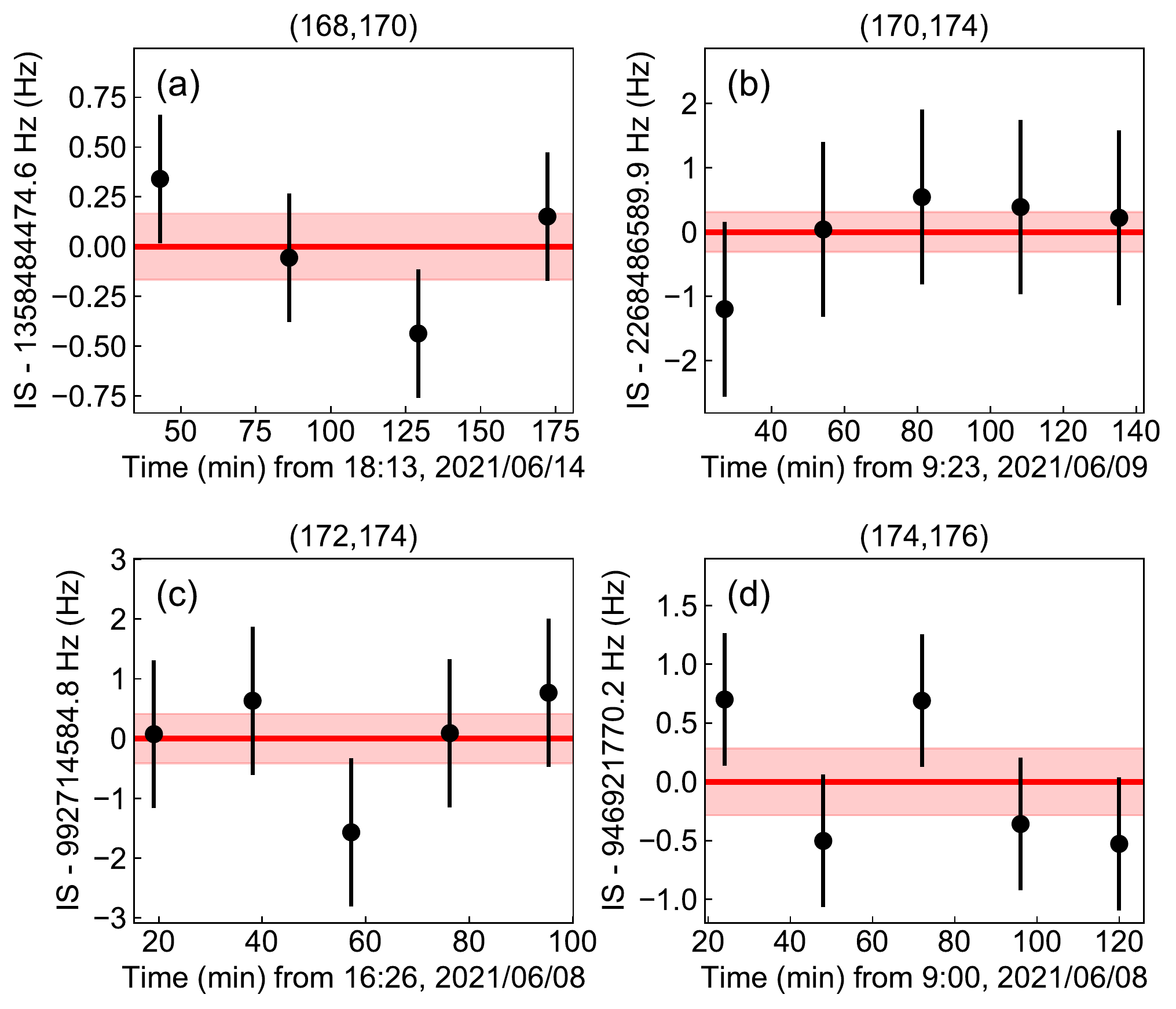}
\caption{Time traces of IS measurements: (a) (168, 170), $\tau=258$~s, $\chi^2_\text{red}=1.03$, (b) (170, 174), $\tau = 162$~s, $\chi^2_\text{red}=0.51$, (c) (172, 174), $\tau=170$~s, $\chi^2_\text{red}=0.75$, and (d) (174, 176), $\tau = 144$~s, $\chi^2_\text{red}=1.13$.  Here, $\chi_\text{red}^2$ represents the reduced chi-squared value. Data points show the mean values of (a) nine measurements, and (b)-(d) ten measurements, and error bars represent the overlapping Allan deviations. Red lines and shaded regions correspond to the means of the data points and the 1$\sigma$ statistical uncertainties, respectively.}
\label{TimeTrace}
\end{figure}

\section{\label{Results}Results}

\subsection{\label{IS measurement}Isotope shifts of the \state{1}{S}{0}--\state{3}{P}{0} transition}
Figures \ref{TimeTrace}(a)-(d) show the results of the IS measurements. We adopt the total lattice depth of $(30, 30, 30)E_\text{R}$ for the operational condition to measure the ISs and their statistical uncertainties. In a typical experiment, fifty measurements are performed, and the data set is divided into five segments. The statistical uncertainty is obtained as the standard error calculated from the mean values of the five segments. 
As a result, the ISs for all four pairs are measured with the statistical uncertainty well below 1~Hz.

\begin{figure}
\includegraphics[width=0.95\linewidth]{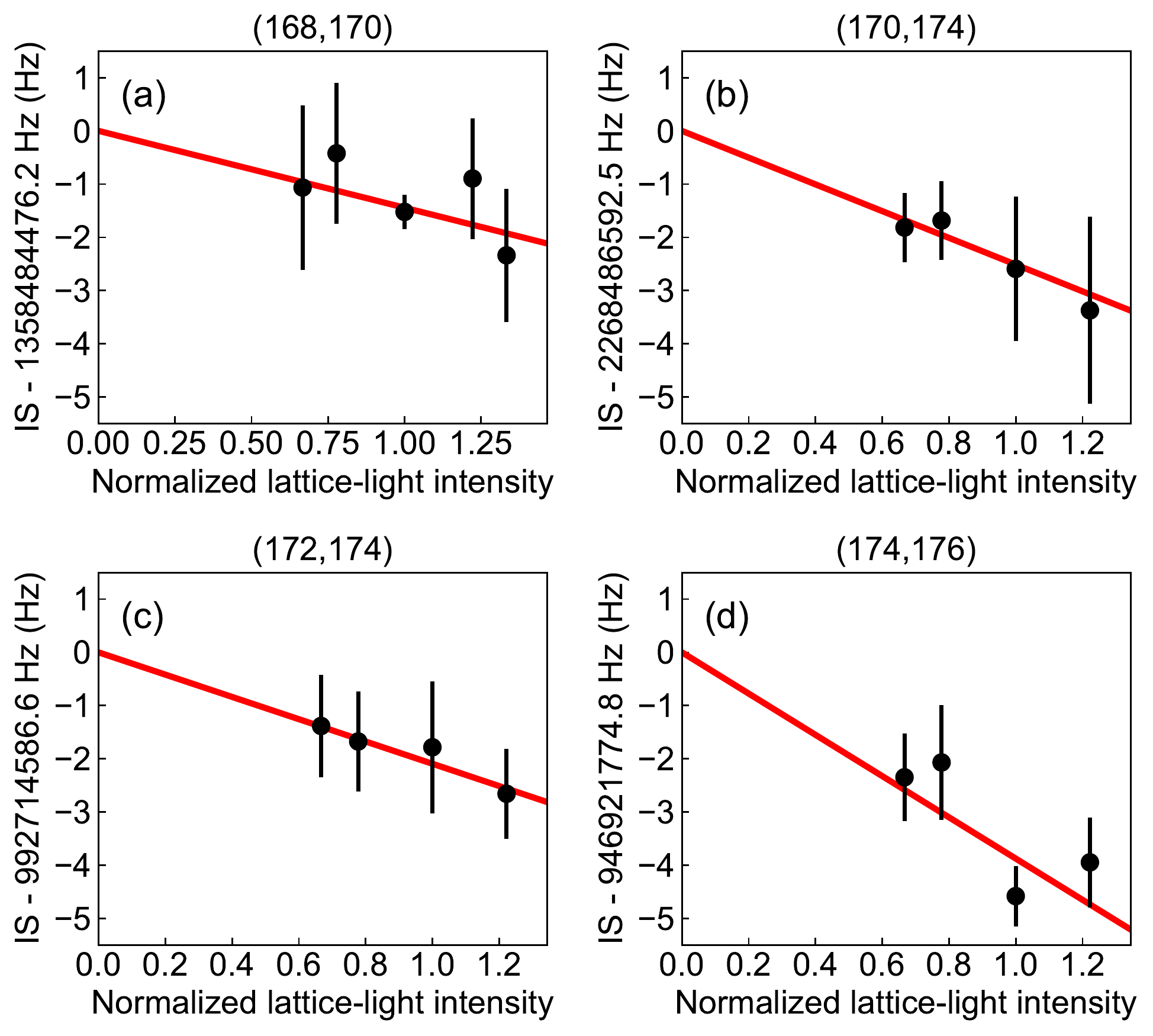}
\caption{Systematic effect of lattice light shift: (a) (168, 170), (b) (170, 174), (c) (172, 174), and (d) (174, 176). The horizontal axis represents the total lattice depth in units of $E_\mathrm{R}$. Error bars show the 1$\sigma$ statistical uncertainties obtained from the overlapping Allan deviations. We draw red lines as guides to the eye. The origin of the vertical axis corresponds to the unperturbed IS, which is obtained from the fit to the data. The inconsistencies between the uncertainties of Figs.~\ref{TimeTrace}(b) and (c) and those in Figs.~\ref{LLS}(b) and (c) at the same operational condition would be ascribed to the small number of measurements in the IS data.}
\label{LLS}
\end{figure}

To compensate for systematic effects on the IS measurements, we experimentally investigate the dependences on the lattice light intensity, the magnetic field, and the probe light intensity. In particular, the lattice light shift is considered to make the predominant contribution since the measurements for all the isotopes are performed at the same lattice wavelength, although, in principle, the magic condition depends on the isotopic species. Figures \ref{LLS}(a)-(d) show the lattice-light-shift dependence of the ISs. These measurements are performed with the $x$-axis lattice depth set to more than $30E_\text{R}$ to satisfy the Lamb-Dicke condition.
The three lattice beams have the different frequencies by the AOMs, and the maximum frequency difference is 160~MHz. According to the differential light shift measured in Ref. \cite{Barber2008}, the maximum light-shift difference between the lattice beams is expected to be less than 0.1~Hz and common for all the isotopes. 
The unperturbed IS is obtained from the fit to the data with the function considering the nonlinear effect due to the zero-point energy \cite{Ushijima2018}.
From this, we determine the correction due to the lattice light shift. The uncertainty in the lattice light shift, which is given by the fitting error, is included as a systematic uncertainty in Table~\ref{SysTable}.

\begin{figure}
\includegraphics[width=0.95\linewidth]{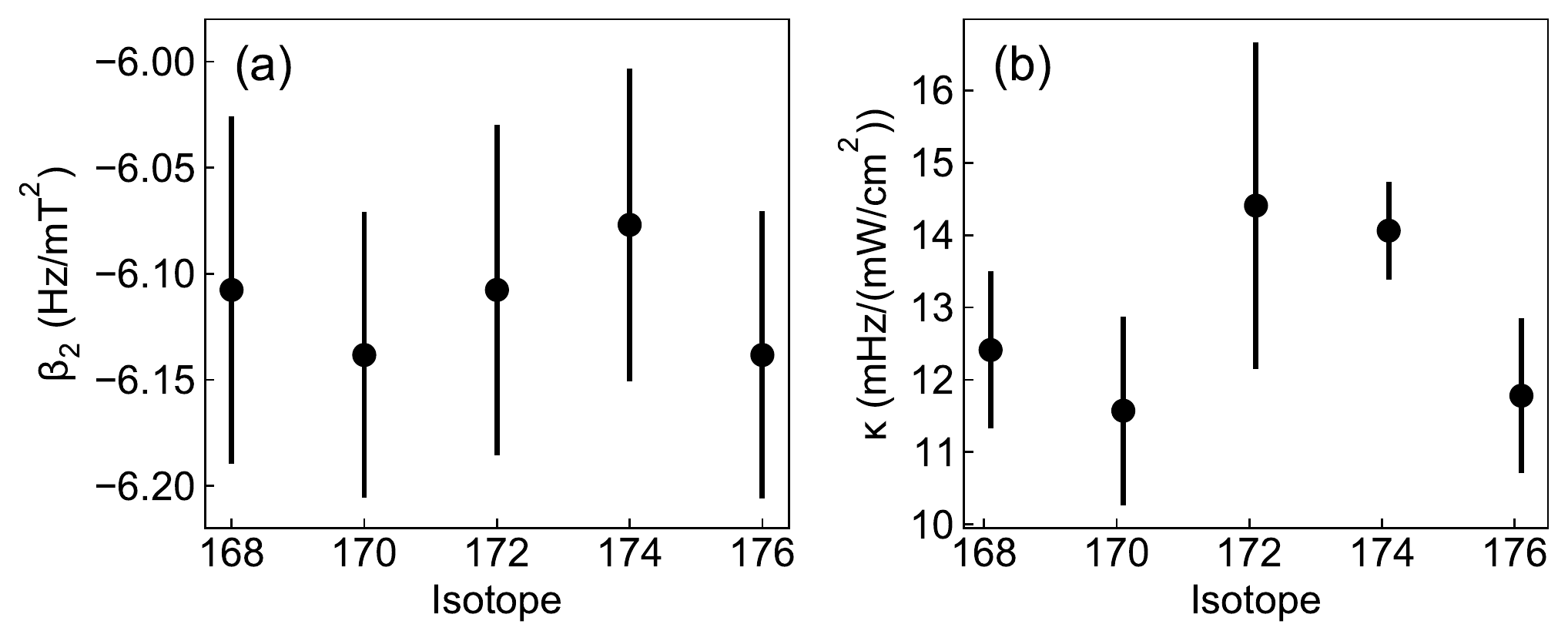}
\caption{Systematic effects of (a) quadratic Zeeman shift, and (b) probe light shift. Error bars show fitting errors with 1$\sigma$ confidence intervals.}
\label{ZSPLS}
\end{figure}

\begin{table*}[tb]
\tabcolsep = 0.2cm
\centering
  \caption{Systematic corrections and uncertainties for measurements of ISs.}
  \begin{tabular}{r r r r r r r r r r} 
   \hline\hline
   \multirow{2}{*}{Systematic effects}&\multicolumn{2}{c}{$(168, 170)$}&\multicolumn{2}{c}{$(170, 174)$}&\multicolumn{2}{c}{$(172, 174)$}&\multicolumn{2}{c}{$(174, 176)$}\\
 & Cor. (Hz) & Unc. (Hz) & Cor. (Hz) & Unc. (Hz) & Cor. (Hz) & Unc. (Hz) & Cor. (Hz) & Unc. (Hz) \\\hline
Lattice light shift & 1.5 & 1.3 & 2.6 & 0.7 & 1.8 & 0.4 & 4.6 & 2.3 \\
Zeeman shift & 0.0 & 0.01 & 0.0 & 0.01 & 0.0 & 0.01 & 0.0 & 0.01 \\
Probe light shift & 0.0 & 1.3 & 0.0 & 1.3 & 0.0 & 1.5 & 0.0 & 1.4 \\
Servo error & 0.1 & 1.1 & 0.1 & 1.2 & 0.0 & 1.4 & 0.1 & 0.9 \\
BBR & 0.0 & 0.02 & 0.0 & 0.02 & 0.0 & 0.02 & 0.0 & 0.02 \\
AOM chirp & 0.0 & 0.01 & 0.0 & 0.01 & 0.0 & 0.01 & 0.0 & 0.01 \\\hline
Total & 1.6 & 2.1 & 2.6 & 1.9 & 1.8 & 2.0 & 4.6 & 2.8 \\
    \hline\hline
\end{tabular}
\label{SysTable}
\end{table*}

Figure \ref{ZSPLS}(a) shows the systematic investigation of the Zeeman shift for each isotope. Since the bosonic isotope has no nuclear spin, there is no linear Zeeman shift in the \state{1}{S}{0} and \state{3}{P}{0} states.
However, the \state{3}{P}{1} state in the fine-structure levels interacts with a magnetic field $B$, yielding the non-negligible quadratic Zeeman shift $-\beta_2 B^2$ in the \state{3}{P}{0} state. To obtain the coefficient $\beta_2$, the interleaved measurement with a target magnetic field and a reference magnetic field $B_0$ is performed, and the magnetic field is calibrated with the \state{1}{S}{0}($F=5/2,m_F=\pm5/2$)--\state{3}{P}{0}($F'=5/2,m_{F'}=\pm5/2$) transitions of \Yb{173}, with the first-order Zeeman shift of $\pm2.768(13)$~Hz/$\mu$T~\cite{Oppong2019}. The mean of the $\beta_2$-values shown in Fig.~\ref{ZSPLS}(a) is $\bar{\beta_2}=-6.11(7)$~Hz/mT$^2$, which is in good agreement with the previous measurement -6.12(10)~Hz/mT$^2$~\cite{Poli2008} and the theoretical value -6.2~Hz/mT$^2$~\cite{Taichenachev2006}. Since the quadratic Zeeman shift originates from the Zeeman mixing between the \state{3}{P}{0} and \state{3}{P}{1} states, it is proportional to the inverse of the energy difference between the two states. 
Therefore, the isotope-dependent Zeeman shift in a magnetic field of $B_0$ is estimated to be $10^{-6}$ times smaller than the observed common shift of 13~Hz at $B_0$. We thus set the correction due to the Zeeman shift to zero, and the systematic uncertainties are given by the uncertainty of the magnetic field, which are summarized in Table~\ref{SysTable}.

Probe light induces a Stark shift $\kappa I$ due to the off-resonant coupling such as the \state{1}{S}{0}--\state{1}{P}{1} transition at 399~nm and the \state{3}{P}{0}--\state{3}{S}{1} transition at 649~nm. Theoretically, the isotope-dependent probe light shift with the intensity $I_0$ is estimated to be $10^{-6}$ times smaller than the observed common shift of 5.6~Hz at $I_0$. As in the measurement of $\beta_2$, the interleaved measurements with target probe intensity $I$ and a reference probe intensity of $I_0$ are performed to obtain the coefficient $\kappa$.
As shown in Fig.~\ref{ZSPLS}(b), the measured probe light shift coefficient $\kappa$ does not show a systematic isotope dependence within the uncertainties of the measurements which are much larger than the expected isotope dependence.
The mean of the $\kappa$-values shown in Fig.~\ref{ZSPLS}(b) is $\bar{\kappa}= 13.3(1.4)$~mHz/(mW/cm$^2$), which is consistent with the previous measurement in Ref.~\cite{Poli2008}.
Similar to the measurement of the quadratic Zeeman effect, the correction due to the probe light shift is set to zero, and the systematic uncertainties are conservatively given by considering the probe-light intensity fluctuation of 20~\%, which are summarized in Table~\ref{SysTable}.

In addition, the systematic effect due to the servo error of the clock operation is considered. The frequency shift from the atomic resonance due to the offset of the error signal is predominantly caused by the drift of the ULE cavity, which is typically 20~mHz/s. The correction and uncertainty due to the servo error are estimated from the mean values and the standard deviations of the excitation fractions on both shoulders of the excitation profile. The results are summarized in Table~\ref{SysTable}. 
Aside from the servo error, the drift of the ULE cavity during the dead time could cause a frequency shift. In our system, the frequency shift by $0\sim1$~Hz is estimated by the fit to the time trace of the EOM frequency during the clock operation. Note that this correction is already taken into account but is not shown in Table~\ref{SysTable} since the frequency drift rate is time-varying and the correction to the data depends on the measurement time.   

The black-body radiation (BBR) shift on the \state{1}{S}{0}--\state{3}{P}{0} transition is calculated as -1.2774(6)~Hz at the temperature of 300~K \cite{Beloy2012}, which is a common perturbation, similar to the Zeeman shift and the probe light shift. Considering the maximum temperature fluctuation in the course of our experiments of 1~K, a change in the BBR shift is estimated as 20~mHz. In addition, the isotope-dependent BBR shift is estimated to be $10^{-6}$ times smaller than the common shift, which leads us to neglect the effect of the BBR shift.

A phase shift that arises when the interrogation light is switched on and off could cause a systematic effect, known as the AOM chirp. 
This is especially serious for a Ramsey resonance.
In our experiment, the switching by the AOM2 in Fig.~\ref{Schematics}(c) may be responsible for this effect.
We examine the phase chirp by mixing down the beat signal between the light passing through the AOM2 and the reference light, yielding a frequency shift of 9(6)~mHz, which is common for all isotopes, and is negligible.

Table~\ref{ISTable} summarizes the measured ISs, showing that each IS is determined with an accuracy of a few~Hz. In addition to bosonic isotope pairs, we measure the IS of \Yb{171} and \Yb{174} to check the accuracy of our measurement scheme by comparing our result with those in the previous studies using the state-of-the-art optical lattice clock technique~\cite{Poli2008, Riehle2018}. Spin-polarized \Yb{171} atoms are prepared instead of the application of the PA, and \Yb{171}($m_F=+1/2$), \Yb{171}($m_F=-1/2$), and \Yb{174} atoms are alternately interrogated by the clock laser. The systematic uncertainty associated with the first-order Zeeman shift of \Yb{171}($m_F=\pm1/2$) is evaluated from $\mp2.000(3)$~Hz/$\mu$T in Ref.~\cite{Ono2018}.
As shown in Table~\ref{ISTable}, our measurement is in good agreement with the results in Refs.~\cite{Poli2008, Riehle2018} with the uncertainty of about 2~Hz.

\renewcommand{\arraystretch}{1.6}
\begin{table}[tb]
\tabcolsep = 0.2cm
\centering
  \caption{Measured ISs of the $\gamma$: \state{1}{S}{0}--\state{3}{P}{0} transition. Total $1\sigma$ uncertainties are shown as $(\vdot)_\mathrm{tot}$.}
  \begin{tabular}{c r c} \hline\hline
 Isotope pair $(A', A)$ & \multicolumn{1}{c}{IS $\nu_\gamma^{A'A}$(Hz)}&References \\ \hline  
(168, 170) & 1358 484 476.2$(2.2)_\mathrm{tot}$&This work \\\hline
(170, 174) & 2268 486 592.6$(1.9)_\mathrm{tot}$&This work \\\hline
(172, 174) & 992 714 586.6$(2.1)_\mathrm{tot}$&This work \\\hline
(174, 176) & 946 921 774.9$(2.9)_\mathrm{tot}$&This work \\\hline
\multirow{2}{*}{(171, 174)} & 1811 281 646.7$(2.3)_\mathrm{tot}$&This work \\
                 & 1811 281 645.8$(0.9)_\mathrm{tot}$&\cite{Poli2008, Riehle2018}  
    \\\hline\hline
  \end{tabular}
\label{ISTable}
\end{table}

\begin{figure}
\includegraphics[width=0.95\linewidth]{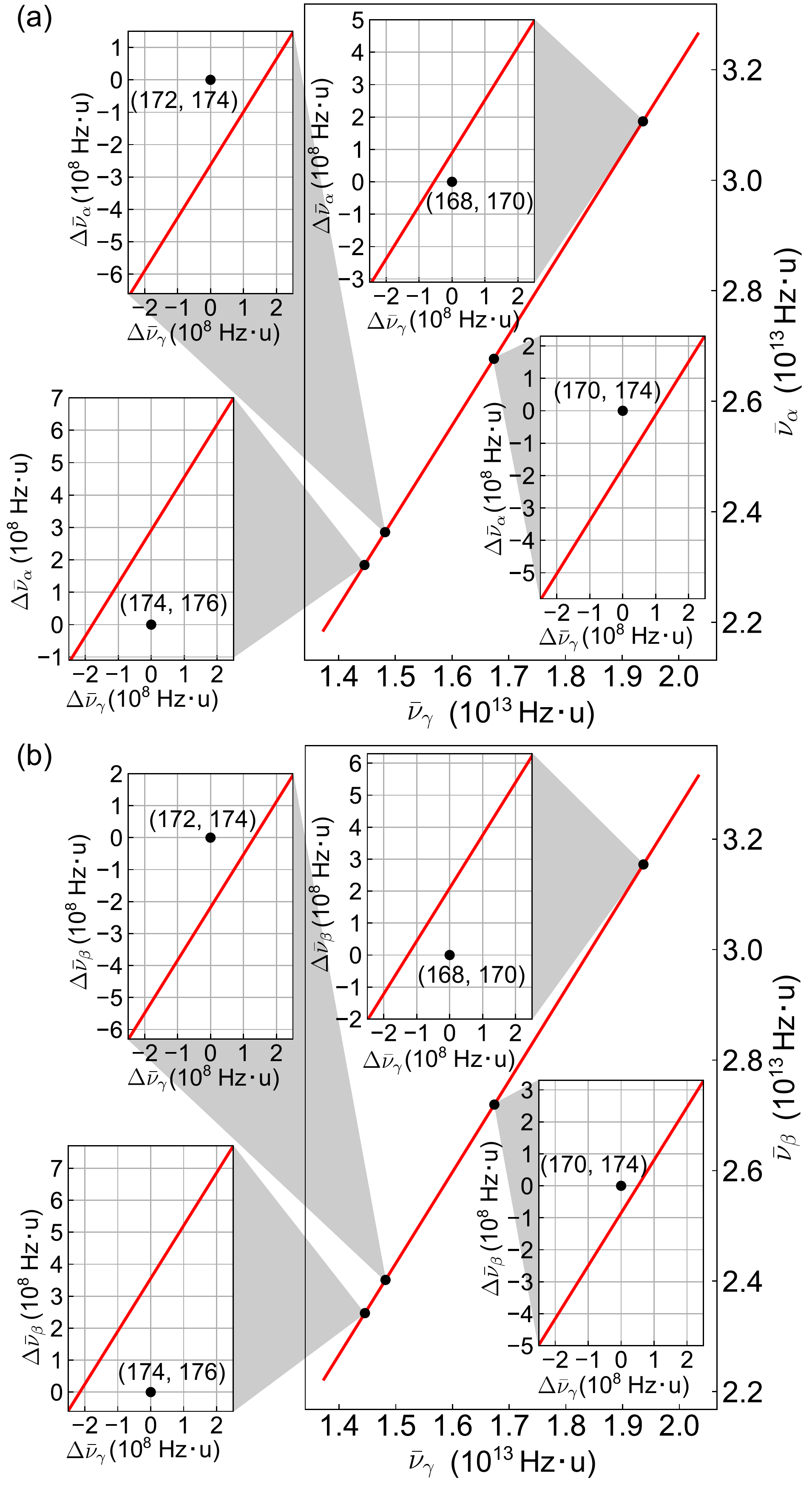}
\caption{2D King plots of $\gamma$: Yb \state{1}{S}{0}--\state{3}{P}{0} transition versus (a) $\alpha$: Yb$^{+}$ \state{2}{S}{1/2}--\state{2}{D}{5/2} transition, and (b) $\beta$: Yb$^{+}$ \state{2}{S}{1/2}--\state{2}{D}{3/2} transition. Each inset shows the zoom-in view magnified by 10$^5$. Error bars, which represent 1$\sigma$ uncertainties, are smaller than the symbol sizes. Solid lines are the fits to the data with Eq.~\eqref{Eq2dKing}.}
\label{2DKP}
\end{figure}

\subsection{\label{2D King plot}2D King plot}
The measured ISs of the $\gamma$: \state{1}{S}{0}--\state{3}{P}{0} can be exploited to study the nonlinearity of the King plot by combining the $\alpha$: \state{2}{S}{1/2}--\state{2}{D}{5/2} and $\beta$: \state{2}{S}{1/2}--\state{2}{D}{3/2} transitions of Yb$^+$ reported in Ref.~\cite{Counts2020}.
Figures \ref{2DKP}(a) and (b) show the 2D King plots with the combinations of $(\bar{\nu}_\gamma^{A'A}, \bar{\nu}_\alpha^{A'A})$ and $(\bar{\nu}_\gamma^{A'A}, \bar{\nu}_\beta^{A'A})$ for the modified IS. Although the results exhibit the overall linearity for both combinations, the deviations on the order of 1$\sim$10~kHz are clearly seen. The fit to the data is obtained by the full $\chi^2$ analysis (see Appendix~\ref{Fitting}), and the best-fit parameters are summarized in Table~\ref{2DKParam}. This is basically the same method used in Refs.~\cite{Mikami2017, Counts2020} except for rearrangements of the data employed in Ref.~\cite{Counts2020}.
The best-fit parameters are roughly consistent with the numerical results in Refs.~\cite{Counts2020, Berengut2020, Schelfhout2021}. 
However, it is also clear that the different theories result in the different values, which are also different from the experimental data. 
The experimental data can be used as a benchmark to improve theoretical calculations.
This analysis also quantifies the nonlinearities as the $\chi^2_{[\la_1,\la_2]}$ minima with the degrees of freedom (dof) of 3,
\begin{align}
  \ch^2_{[\ga,\al]}=& 1.1\times 10^4, 
  \label{chi2ca}\\
	\ch^2_{[\ga,\be]}=& 1.7\times 10^4,
 \label{chi2cb}
\end{align}
which are much larger compared to that of the transitions $(\alpha, \beta):\ch^2_{[\al,\be]}=15.37$ (see Appendix~\ref{2DKP of AB}). The observed large nonlinearity compared with that of the transitions $(\al, \be)$ is reasonable when we consider the characters of the transitions involved. 
The transitions $\al$ and $\be$ are the same with each other except for the relativistic effects, which results in the almost unity for the value of $F_{\be\al}$ and thus $H_{\be\al}\sim H_\be-H_\al \sim 0$ when $H_\al \sim H_\be$.
On the other hand, the electronic configuration of the transition $\ga$ is quite different from those of $\al$ and $\be$.
As a result, $F_{\al\ga}$ and $F_{\be\ga}$ take about 1.6, different from unity, and thus there is no cancellation in $H_{\al\ga}$($H_{\be\ga}$) even when $H_\ga \sim H_\al$ ($H_\ga \sim H_\be$).

We introduce the QFS 
as the higher-order IS to fit the data. 
We take $[\delta\langle r^2\rangle^2]^{A'A}$ as $\delta\eta^{A'A}$ in Eq.~\eqref{Eq2d_NL} and its coefficient as an additional fitting parameter (see Appendix~\ref{Fitting} for details).
However, $\ch^2$ is not improved,   
\begin{align}
  \ch^2_{[\ga,\al]}(\text{QFS}) =& 1.0\times 10^4 ,\\
  \ch^2_{[\ga,\be]}(\text{QFS}) =& 8.4\times 10^3,
\end{align}
where the dof of $\chi^2$ is 2.
This indicates several possibilities.
While we use the mean-square nuclear charge radii shown in Table~\ref{TabMass} in the evaluation of QFS term, these values may not be accurate enough at the present level of experimental accuracy, which may cause the fitting worse.
The failure of the fitting can be also explained when the origin of the observed nonlinearity cannot be solely attributed to the QFS.
In fact, the roles of the next leading order Seltzer moment $\de\avg{r^4}^{A'A}$~\cite{Tanaka2019} and the nuclear deformation~\cite{Allehabi2021} are discussed.
The assumption of the approximate relation of $\de\avg{r^4}^{A'A}$ to $[\de\avg{r^2}^2]^{A'A} $, which is the basis of the absorption of the $\de\avg{r^4}^{A'A}$ term into the QFS and leading order FS term \cite{Counts2020}, may not be validated
at the improved accuracy of the measurements associated with the transition $\ga$. 
 
We also introduce the PS as the higher-order IS to fit the data. 
However, again $\ch^2$ is not improved,   
\begin{align}
  \ch^2_{[\ga,\al]}(\text{PS})   =& 8.8\times 10^3 ,\\
  \ch^2_{[\ga,\be]}(\text{PS})   =& 7.5\times 10^3 .
\end{align}

\renewcommand{\arraystretch}{1.6}
\begin{table}[tb]
\tabcolsep = 0.2cm
\centering
  \caption{Best-fit parameters of 2D King plots shown in Figs.~\ref{2DKP}(a) and (b). Here $\la_1$ represents the optical transition of $\ga$, and $\la_2$ $\al$ or $\be$. The dof of $\chi^2$ is 3 = 5(observations for $\al$ or $\be$)+4(observations for $\ga$)-2(fitting parameters)-4(ISs on $\ga$). The error of each fit parameter is evaluated as $1 \si$. The results are compared with the theoretical results in Refs.~\cite{Counts2020, Berengut2020, Schelfhout2021}. The values associated with Refs~\cite{Berengut2020, Schelfhout2021}(Refs.~\cite{Counts2020,Schelfhout2021}) are obtained by combining the MS and the FS of the $\ga$ transition in Ref.~\cite{Schelfhout2021} and those of the $\al$ and $\be$ transitions in Ref.~\cite{Berengut2020}(Ref.~\cite{Counts2020}).}
  
  \begin{tabular}{c r r c} \hline\hline
 $(\lambda_1,\lambda_2)$ & \multicolumn{1}{c}{$F_{\la_2\la_1}$}& \multicolumn{1}{c}{$K_{\la_2\la_1}$~(GHz$\vdot$u)} &References \\ \hline  
\multirow{3}{*}{$(\ga, \al)$} & $1.63471003(90)$ & $-593.388(16)$ &This work \\
                              & $1.5140        $ & $-255        $ & \cite{Berengut2020} \\
                              & $1.3565        $ & $-856        $ & \cite{Berengut2020, Schelfhout2021}\\
                              & $1.4613        $ & $-1257       $ & \cite{Counts2020, Schelfhout2021}\\\hline
\multirow{3}{*}{$(\ga, \be)$} & $1.6533771(10) $ & $-480.159(16)$ &This work \\
                              & $1.5400        $ & $-31         $ &\cite{Berengut2020}  \\ 
                              & $1.3798        $ & $-643        $ &\cite{Berengut2020, Schelfhout2021} \\
                              & $1.4836        $ & $-1211       $ &\cite{Counts2020, Schelfhout2021} \\
                                           \hline\hline
  \end{tabular}
\label{2DKParam}
\end{table}

\begin{figure}
\includegraphics[width=0.95\linewidth]{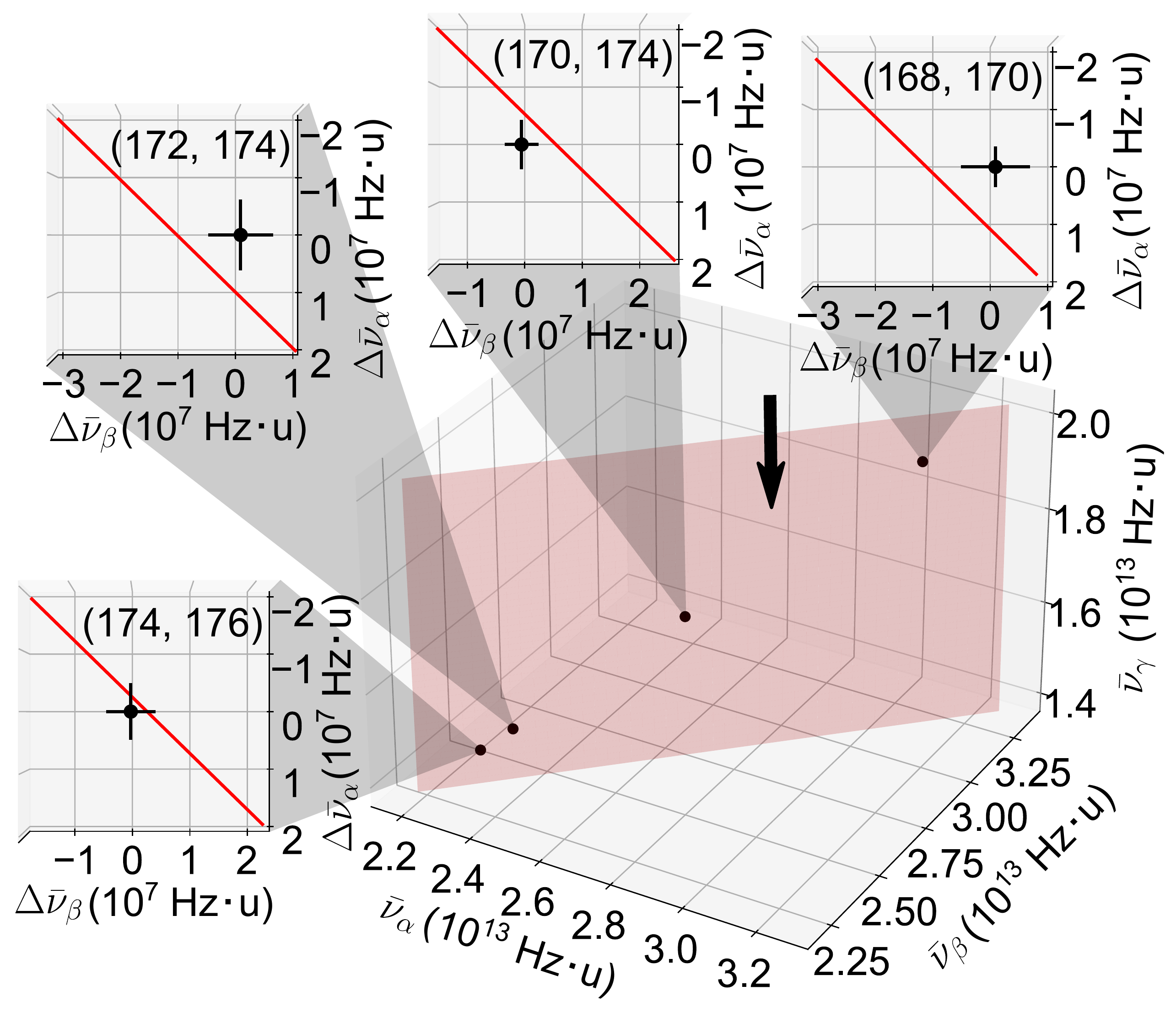}
\caption{Generalized King plot using all three transitions. The plane is the fit to the data by Eq.~\eqref{Eq3dKing_mIS}. Each inset shows the zoom-in view magnified by $10^6$, seen from the direction parallel to the plane and orthogonal to the $\be$ axis (indicated by the arrow), and so the plane is represented as the red line in the inset. The labels $\Delta\bar{\nu}_\alpha$ and $\Delta\bar{\nu}_\beta$ show the deviations from the modified ISs, and a label for the $\gamma$ axis is not shown in the insets. Error bars represent 1$\sigma$ uncertainties.}
\label{3DKP}
\end{figure}
\subsection{\label{3D King plot}3D King plot}
\subsubsection{\label{nonlinearity of 3D King plot}Observation of nonlinearity of 3D King plot}
The generalized King plot combining all the IS data for the three transitions provides us with important insights on the origin of the violation of the King linearities Eqs.~\eqref{chi2ca} and~\eqref{chi2cb}.  
As we discuss in Sec.~\ref{Isotope shifts and 3D King plot}, if the observed nonlinearity originates from the single source which is factorized as the products of isotope-dependent nuclear term and isotope-independent electronic term, we expect the perfect linearity of the generalized 3D King plot Eq.~\eqref{Eq3dKing_mIS}. 
Figure~\ref{3DKP} shows the generalized King plot with the three transitions $(\la_1,\la_2,\la_3)=(\ga,\al,\be)$, and the best-fit parameters are summarized in Table~\ref{3DKPParam}.
Note that the $\abs{f_\ga}$ is much smaller than $f_\al\approx1$, indicating the best-fit plane is almost perpendicular to the horizontal, or $\al$--$\be$ plane, as shown in Fig.~\ref{3DKP}. This is reasonable if we consider the close similarity between the $\al$ and $\be$ transitions and much less between the $\ga$ and these two. The experimental data can be used as a benchmark to improve theoretical calculations.
The minimum $\ch^2$ of this linear relation is
\begin{align}
  \ch^2_{[\ga,\al,\be]} = 15 \quad (p=2.3\times 10^{-3}),
  \label{chi2abc}
\end{align}
where the corresponding $p$-value is calculated with the dof of 3. This $p$-value corresponds to a significance of 3$\sigma$.
In Sec.~\ref{2D King plot} of 2D King plots including the transition $\ga$, we have seen that the $\ch^2$ minima are quite large.
In the generalized 3D King plot, on the other hand, we find that $\ch^2$ is small, which indicates that the major origin of the nonlinearity is removed.

\renewcommand{\arraystretch}{1.6}
\begin{table}[tb]
\tabcolsep = 0.2cm
\centering
  \caption{Best-fit parameters of 3D King plots shown in Fig.~\ref{3DKP}. The error of each fit parameter is evaluated as $1 \si$. The results are compared with the theoretical results Ref.~\cite{Berengut2020} under the assumption that the QFS is eliminated by the construction of the 3D King plot.}
  
\begin{tabular}{rrrc} \hline\hline
 \multicolumn{1}{c}{$f_{\al}$}&\multicolumn{1}{c}{$f_{\ga}$}& \multicolumn{1}{c}{$k_\mu$~(GHz$\vdot$u)} & Reference \\ \hline  
                                       $1.023(13)$ & $-0.019(21)$ & $127.2(7.7)$ & This work\\ \hline
                                       $1.276$     & $-0.391$     & $294.4$     & \cite{Berengut2020} \\
 \hline\hline
  \end{tabular}
\label{3DKPParam}
\end{table}

%
%


%
Importantly, at the same time, the obtained $\ch^2$ is not as small as the expected minimum value $\chi^2_\text{min}=4.037$ limited by the consistency condition $\nu_\la^{A'A} =\nu_\la^{A'A_0}-\nu_\la^{A A_0}$ for the $\al$ and $\be$ transitions (see Appendix~\ref{2DKP of AB}).  
This means that the observed ISs data cannot be explained by a single source of the higher-order effects such as the QFS or the nuclear deformation proposed as candidates of the observed nonlinearity in the $(\al, \be)$ King plot~\cite{Counts2020, Allehabi2021}, or the PS, but at least two distinct higher-order ISs are involved in the data.

\subsubsection{\label{MIT analysis}Inapplicability of the PS and QFS assumption}
In the derivation of the bound on the new boson in Refs.~\cite{Counts2020,Berengut2020}, the PS and only the QFS within the SM are assumed to be the two distinct higher-order ISs involved in the data of Ref~\cite{Counts2020}.
In the Appendix~\ref{finite value}, we describe the result of a 3D analysis under the same assumption.
Note that the purpose of this analysis is just to check the validity of this assumption adopted in Refs.~\cite{Counts2020,Berengut2020}.
This analysis results in the inconsistently large positive signal for the new boson, which demonstrates that the above assumption which considers only one SM effect of QFS  and neglects all other SM effects is not justified.

\subsubsection{\label{upper-bound}Determination of the upper bound of the new particle coupling}
Here, we present the main result of this work, namely, we set the upper bound of the new particle coupling from the analysis of our 3D King plot. We assume that there is an additional SM source of the higher-order IS other than the QFS, and this additional SM source is to be eliminated by the construction of the 3D King plot as discussed in Sec.~\ref{Isotope shifts and 3D King plot}.
In addition, we assume that the remaining nonlinearity observed in the 3D King plot is originated from the QFS. 
Here, we make argument on the validity of the assumptions in the above. As shown in \ref{3D King plot}, the fit with the original 3D King’s relation Eq.~\eqref{Eq3dKing_mIS} exhibits the nonlinearity at the 3$\sigma$ level, indicating that there exist at least two distinct higher-order contributions to the measured isotope shifts. It is reasonable that the two distinct higher-order IS terms involved in the observed data are the next-leading order Seltzer moment $\delta\langle r^4\rangle^{A'A}$ and the QFS (see Refs. \cite{Counts2020, Blundell1987, Flambaum2018, Tanaka2019, Allehabi2021} and Eq.~\eqref{EqNonlinear}) since these two terms are the next-order corrections for the leading order field shift. Note that the higher-order terms in the mass shift are much smaller than these corrections for heavy elements such as Yb \cite{Flambaum2018, Palmer1987}. Although the source of nonlinearity other than QFS is not needed to be specified in our analysis of the 3D King plot, the next-leading order Seltzer moment $\delta\langle r^4\rangle^{A'A}$ is then a plausible candidate for another source.

\begin{figure}
\includegraphics[width=0.95\linewidth]{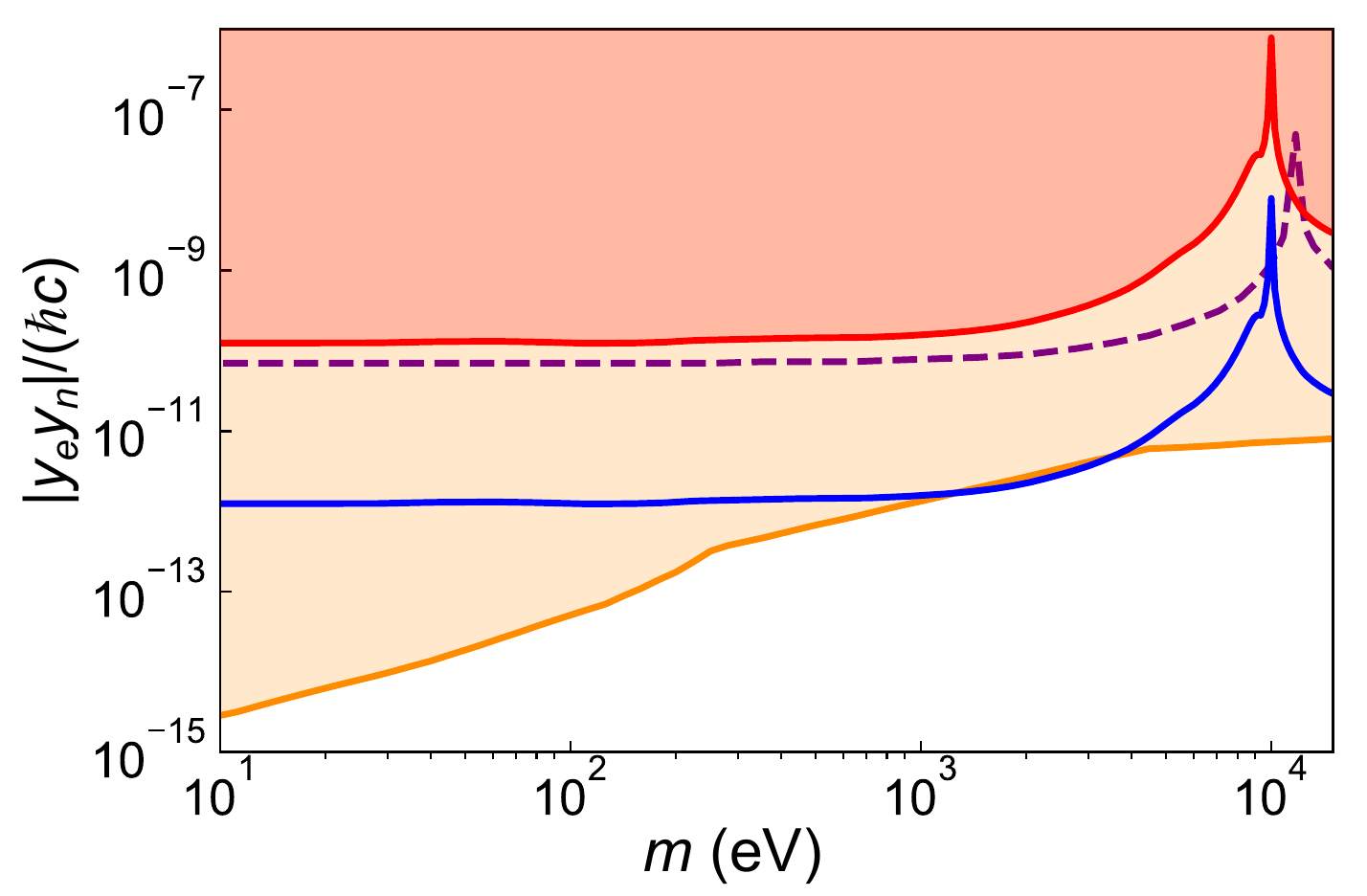}
\caption{Product of couplings $|y_ey_n|$ of a new boson as a function of the mass $m$. The red line represents the upper bound of the coupling with the 95\% C.L. (see main text for details). For comparison, the constraint from electron anomalous magnetic moment $(g-2)$ measurements \cite{Hanneke2008, Morel2020} combined with neutron scattering measurements \cite{Leeb1992,Pokotilovski2006,Nesvizhevsky2008} is shown as the orange shaded region. The purple dashed line shows the upper bound from the Ca$^+$ analysis in Ref.~\cite{Solaro2020}. The blue line shows the constraint which is obtained if the uncertainties of $\nu_{\al(\be)}^{A'A}$ and $\de\avg{r^2}^{A'A}$ are improved by factors of $10^2$ and 10, respectively.}
\label{NPBound}
\end{figure}

Under these assumptions, we perform the QFS fit, namely the fit with Eq.~\eqref{Eq3d_NL} with $h=h_\text{QFS}$ and $\delta\eta^{A'A}=[\delta\avg{r^2}^2]^{A'A}$.
The result of the fitting shows that the $\chi^2$ minimum reaches the theoretical minimum.
This procedure gives a constraint on $h_\text{QFS}$ and we take this constraint into account in the following $\chi^2$ analysis. 
Then, we additionally introduce the nonlinear term given by $h_\text{PS}$ as in Eq.~\eqref{Eq3d_NL}, and evaluate the upper bound of $h_\text{PS}$ as the value with which the $\chi^2$ increases from the $\chi^2$ minimum by the amount corresponding to the 95$\%$ C.L.
The electronic factors $X_{\al}$, $X_{\be}$, and $X_{\ga}$ in the expression of $h_\text{PS}$ are similarly taken from Refs.~\cite{Berengut2020,Counts2020} as in Appendix~\ref{finite value}, while the coefficients $f_{\al}$ and $f_{\ga}$ in $h_\text{PS}$ are fixed as the best-fit values obtained from the fitting as the coefficients of $\nu_{\al}^{A'A}$ and $\nu_{\ga}^{A'A}$ in Eq.~\eqref{Eq3d_NL}.
The adoption of the values of $f_{\al}$ and $f_{\ga}$ is reasonable since these should coincide with the coefficients appearing in the $h_\text{PS}$ (see Appendix~\ref{Formulation of h}).
The obtained upper bound is shown as the red curve in Fig.~\ref{NPBound}.
If the mass of the new boson is 10~eV, we obtain the bound of $|y_e y_n|/(\hbar c) \leq 1.2\times 10^{-10}$ with the 95\% C.L.
This bound is of the same order as the one with the data of Ca$^+$ \cite{Solaro2020}, which is $|y_ey_n|/(\hbar c)<6.9\times{10}^{-11}$ for $m=1$~eV and depicted as the purple dashed line in Fig.~\ref{NPBound}. 
While this is above the terrestrial bound given as the orange line in Fig.~\ref{NPBound}, the sensitivity obtained from the 3D King plot can surpass this terrestrial bound for $m\sim2$~keV if the uncertainties of $\nu_{\al(\be)}^{A'A}$ and $\de\avg{r^2}^{A'A}$ are improved by factors of $10^2$ and 10, respectively (see the blue line shown in Fig.~\ref{NPBound}).
The uncertainty in the nuclear mass coming from those of atomic mass and  binding energy does not play an important role in the $\chi^2$ value, nor needs to be improved for the blue line.
We note that the above upper bound is applicable provided no accidental cancellation between the QFS and PS contributions to the observed nonlinearity occurs in the 3D King plot.

In general, it is difficult to discriminate the origin of the higher-order IS terms among the different sources. The obtained fitting coefficients $f_\al$, $f_\ga$, $k_\mu$ will offer a possible consistency check in the nonlinearity fit (see Appendix~\ref{Consistency check}).

\subsection{\label{Absolute frequency}Absolute frequency of clock transition}
In addition, by referencing the reported absolute frequency of the  \state{1}{S}{0}--\state{3}{P}{0} transition for \Yb{171} \cite{Riehle2018}, our IS measurements provide the absolute frequencies for five bosonic isotopes, which is summarized in Table~\ref{AbsFreq}.

\renewcommand{\arraystretch}{1.6}
\begin{table}[tb]
\tabcolsep = 0.2cm
\centering
 \caption{Absolute frequencies of $\gamma$: \state{1}{S}{0}--\state{3}{P}{0} transition.}

\begin{tabular}{clc}
\hline \hline
 Isotope &  \multicolumn{1}{c}{Transition frequency (Hz)}  & References \\
 \hline
168 & 518 297 652 280 285.8(3.7)& This work  \\\hline
170 & 518 296 293 795 809.6(3.0)& This work  \\\hline
171 &518 295 836 590 863.6(0.3)& \cite{Riehle2018} \\\hline
172 & 518 295 018 023 803.6(3.1)& This work  \\\hline
173 & 518 294 576 845 268(10)&  \cite{Clivati2016} \\\hline
\multirow{2}{*}{174}  & 518 294 025 309 216.9(2.3)&This work \\
 & 518 294 025 309 217.8(0.9) &\cite{Poli2008}\\\hline
176 &518 293 078 387 442.1(3.7)&This work \\
\hline \hline
\end{tabular}
\label{AbsFreq}
\end{table}

\section{\label{Conclusion}Conclusions and prospects}
In conclusion, we measure ISs for neutral Yb isotopes on an ultranarrow optical clock transition \state{1}{S}{0}--\state{3}{P}{0} with an accuracy of a few Hz. 
The determined ISs are combined with the recently reported IS measurements for two optical transitions of Yb$^+$, enabling us to construct the two King plots. 
Both of them show very large nonlinearity, demonstrating the high sensitivity to the higher-order effect in the IS.  
We also carry out the generalized King plot using the three optical transitions.
Our analysis shows a deviation from linearity at the 3$\sigma$ uncertainty level and $|y_e y_n|/(\hbar c) < 1\times 10^{-10}$ for $m<1$~keV with the 95\% C.L.

We expect that the demonstrated method for the precise IS measurement will be straightforwardly applied to additional ultranarrow optical transitions of (6s)$^{2}$ \state{1}{S}{0}--6s6p \state{3}{P}{2}  at the wavelength of 507~nm and (4f)$^{14}$(6s)$^{2}$ \state{1}{S}{0}--(4f)$^{13}$(6s)$^{2}$5d $(J=2)$ at 431~nm of Yb atoms~\cite{Safronova2019,Dzuba2018}, providing the original 2D and generalized 3D King plots using neutral Yb transitions alone.
Note that the high-resolution spectroscopy and high-sensitive detection methods for the (6s)$^{2}$ \state{1}{S}{0}--6s6p \state{3}{P}{2} transition are already demonstrated~\cite{Yamaguchi2008, Kato2016, Nakamura2019}.
Furthermore, as we measure the fermionic isotope of \Yb{171}, we can also extend the measurement to another fermionic isotope of \Yb{173}, which will allow us to construct a still higher-dimensional King plot~\cite{Schelfhout2021}.
These will give important insights into the origin of the observed nonlinearity in this work.
Finally, we note that our present work and future efforts offer important benchmarks for studies to discriminate between different nuclear models through theory-experiment comparisons~\cite{Allehabi2021}.

\appendix
\section{\label{mass and radius}Nuclear mass and nuclear charge radius}
As the inverse mass differences, we use the masses of the nuclei.
They are calculated by the formula $m_A = m_\text{nucleus} =m_\text{atom}(A) -Z m_e +E_b(Z)$ with the atomic masses $m_\text{atom}(A)$ given in Table~\ref{TabMass}, where the electron mass $m_e= 5.48 579 909 065(16)\times 10^{-4}\text{u}$, and u represents the atomic mass unit \cite{Huang2021, Wang2021}.
We evaluate the binding energy with $E_b(Z)=14.4381 Z^{2.39}+1.55468\times 10^{-6} Z^{5.35} \text{eV}$ with the uncertainty of 1.1~keV given by Ref.~\cite{Lunney2003}.
In the case of Yb, this is $E_b(70) = 4.10 6(12) \times 10^{-4}\text{u}$ with the conversion factor of $1 \text{eV} =1.073 544 102 33(32)\times 10^{-9} \text{u}$.

The ionic masses are used in Ref.~\cite{Counts2020}. As shown in Appendix~\ref{2DKP of AB}, our 2D analysis of the Yb$^+$ data with the nuclear masses is consistent with Ref.~\cite{Counts2020}. In our other 2D and 3D analyses, the changes in $\chi^2$ are a few percent when the atomic masses are employed instead of the nuclear masses. Thus, the precise definition of masses to describe the first-order MS is not relevant in the present work. In particular, the bound on the new particle and its future prospect given in Fig.~\ref{NPBound} vary only about one percent even if we use the atomic masses.

To evaluate the isotope dependence of the QFS, we use the mean square charge radii given by Ref.~\cite{Schelfhout2021}.
These values are summarized in Table~\ref{TabMass}.

\renewcommand{\arraystretch}{1.6}
\begin{table*}[htb]
\tabcolsep = 0.2cm
\centering
 \caption{
   Atomic masses and differences in mean-square nuclear charge radii. These values are respectively given by Refs.~\cite{Wang2021} and \cite{Schelfhout2021}.
 }
\begin{tabular}{cc}
\hline \hline
  $A$ &  $m_\text{atom}(A)$ (u) \\ \hline
  168 & 167.933 891 30(10)  \\ 
  170 & 169.934 767 243(11) \\
	172 & 171.936 386 654(15) \\
	174 & 173.938 867 546(12) \\
	176 & 175.942 574 706(16) \\
\hline \hline
\end{tabular}
  \hspace{5em}
\begin{tabular}{cc}
\hline \hline
  $(A',A)$ & $\de\avg{r^2}^{A'A} (\text{fm}^2)$ \\ \hline
  (170, 168) & 0.1348(6) \\
	(172, 170) & 0.1266(6) \\
	(174, 172) & 0.0989(6) \\
	(176, 174) & 0.0944(5) \\
\hline \hline
\end{tabular}
\label{TabMass}
\end{table*}

\section{\label{Experimental parameters}Isotope-dependent experimental parameters}

Table \ref{ExpParam} summarizes the experimental parameters which depend on the isotopic species. The PA lines are identified experimentally, whose resonance frequencies are consistent with the theoretical calculation in Ref.~\cite{Borkowski2009}. Instead of the PA, the spin-polarized \Yb{171} atoms are prepared using the optical pumping associated with the \state{1}{S}{0}($F=1/2$)--\state{3}{P}{1}($F'=1/2$) transition. As well as \Yb{171}, the resonant light with \state{1}{S}{0}($F=5/2$)--\state{3}{P}{1}($F'=5/2$) transition is applied to pump the \Yb{173} atoms into the $\ket{\uparrow(\downarrow)}=\ket{m_F=+(-)5/2}$ states. It is noted that the PA light is not applied to the \Yb{173} atoms since the clock transition frequency associated with the multiply occupied sites is well separated from that of the singly occupied sites \cite{Scazza2014, Cappellini2014}. 

\renewcommand{\arraystretch}{1.6}
\begin{table*}[htb]
\tabcolsep = 0.2cm
\centering
 \caption{Summary of isotope-dependent experimental parameters. The parameters $\Delta_A$ and $\tau_A$ correspond to the detuning of the PA line from the \state{1}{S}{0}--\state{3}{P}{1} transition and the PA time for the isotope $A$, respectively. In addition, $N_A$ and $T_A$ correspond to the number of the atom before the interrogation and the atom temperature after the evaporative cooling, respectively. It is noted that the \Yb{174} atoms are cooled below the transition temperature of the Bose-Einstein condensation (BEC) with no discernable thermal components.}

\begin{tabular}{crrrrrr}
\hline \hline
$(A', A)$ & (170, 168) & (174, 170) & (174, 172) & (171, 174) & (174, 176) & ($173\uparrow, 173\downarrow$) \\
\hline
$\tau_0$ (s) & 64.6 & 40.5 & 28.6 & 39.7 & 36 & 34 \\
$\tau$ (s) & 258.4 & 162 & 114.4 & 238.2 & 144 & 136 \\
\hline \hline
\end{tabular}
\vskip0.3cm
\begin{tabular}{crrrrrrr}
\hline \hline
$A$ &168 &170& 171 & 172 & 173 & 174&176 \\
\hline
$\Delta_A/2\pi$ (MHz) & -2072 & -6213 & & -1143 &  &  -3687 & -598  \\
$\tau_A$ (ms) & 10 & 20 & & 10 &  & 1 &1  \\
$N_A$ ($\times10^3$) & 10 & 15 & 10 & 15&20 & 25 & 25 \\
$T_{A}$ ($\mu$K) & 0.5 & 0.3 & 0.5 & 0.7& 0.1 & BEC & 0.2 \\
\hline \hline
\end{tabular}
\label{ExpParam}
\end{table*}

\section{\label{Fitting}Statistical test of linearities}
The nonlinearity of the King plot is analyzed with the $\chi^2$ function. We measured $\nu_\ga^{AA'}$ for the isotope pairs $\mathcal{A}_\ga=\{(168,170), (170, 174), (172, 174),(174, 176)\}$. In Ref.~\cite{Counts2020}, for the transitions $\al$ and $\be$, the isotope pairs $\mathcal{A}_\al=\mathcal{A}_\be=\{(168,170), (170, 172), (172, 174),(174, 176)\}$ are measured. In addition to these pairs, the $\nu_\al^{170,174}$ and $\nu_\be^{170,174}$ are measured in Ref.~\cite{Counts2020}, and we perform the $\chi^2$ analysis including these redundant data. 
Although it is possible to quantify nonlinearities by introducing a geometrical measure like areas of triangles as discussed in Ref.~\cite{Berengut2020}, the $\chi^2$ analysis is more straightforward to handle the redundant data and multiple sources of nonlinearities.

\paragraph{2D case} 
We consider the following model for the 2D King plot with the $(\la_1,\la_2)$ transitions
\begin{equation}
\nu_{\la_2}^{AA'} = c_\mu\de\mu^{AA'} + c_{\la_1}\nu_{\la_1}^{AA'},
\label{2Dfit}
\end{equation}
where $c_\mu$ and $c_{\la_1}$ are the model parameters associated with the electronic factors.
The corresponding $\chi^2$ function is given by
\begin{align}
\chi^2=&\chi^2_\text{mass}
        +\sum_{\la=\la_1,\la_2}\left\{\sum_{(A,A')\in\mathcal{A_\la}}\left(\frac{\nu^{AA'}_{\la}-\tilde{\nu}^{AA'}_{\la}}{\sigma_{\nu^{AA'}_{\la}}}\right)^2\right.\nonumber\\
&\left.+\left(\frac{\nu^{170,172}_{\la}+\nu^{172,174}_{\la}-\tilde{\nu}^{170,174}_{\la}}{\sigma_{\nu^{170,174}_{\la}}}\right)^2\right\},
\label{chi2D}
\end{align}
where the tilded quantities and $\sigma_{(\cdot)}$ correspond to the measured quantities and their experimental uncertainties, respectively. 
The first term $\chi_{\mathrm{mass}}^2$ contains the contributions from the parameters related to nuclear masses described in Appendix~\ref{mass and radius}.
Note that the last term is only included for the transitions $\al$ and $\be$.
For example, the $\chi^2_{[\al,\be]}$, the $\chi^2$ function for $(\la_1,\la_2)=(\al,\be)$ has the following parameters:
four independent $\nu^{AA'}_{\al}$'s,
and two model parameters ($c_\mu$ and $c_{\al}$).
Thus, we have $4+2=6$ fitting parameters.
While, we have the following measurements with errors: 
five $\tilde{\nu}^{AA'}_{\al}$'s and five $\tilde{\nu}^{AA'}_{\be}$'s.
Thus, we have $5+5=10$ experimental constraints.
Hence, the dof is $10-6=4$.
It is noted that the dof of the $\chi^2_{[\al,\be]}$ in Ref.~\cite{Counts2020} is 2 since the redundant measurements $\tilde{\nu}^{170,174}_{\al(\be)}$ are used to improve the precision of $\tilde{\nu}^{170,172}_{\al(\be)}$ and $\tilde{\nu}^{172,174}_{\al(\be)}$, meaning that $\tilde{\nu}^{170,172}_{\al(\be)}$, $\tilde{\nu}^{172,174}_{\al(\be)}$ and $\tilde{\nu}^{170,174}_{\al(\be)}$ are not independent of each other. From the same argument, the dof of the $\chi^2_{[\al,\ga]}$ and $\chi^2_{[\be,\ga]}$ is 3 in our analysis.

When the QFS(PS) is considered as the nonlinear source of the King relation, the fitting function is modified by adding $c_q[\de\avg{r^2}^2]^{AA'}$($c_p(A-A')$) to the right hand side of Eq.~\eqref{2Dfit}, where $c_q$($c_p$) is a fitting parameter for the QFS(PS), and the $\chi^2$ function has the dof of 3 for the case of $(\al,\be)$, and 2 for both of $(\al,\ga)$ and $(\be,\ga)$.

\paragraph{3D case}
As well as the 2D case, the linear model for the 3D King plot with the $(\la_1,\la_2, \la_3)$ transitions is considered
\begin{equation}
\nu_{\la_3}^{AA'} = c_\mu\de\mu^{AA'} + c_{\la_1}\nu_{\la_1}^{AA'}+ c_{\la_2}\nu_{\la_2}^{AA'}.
\label{3Dfit}
\end{equation}
The corresponding $\chi^2$ function has the same form as Eq.~\eqref{chi2D} where the summation for $\lambda$ now includes $\lambda_3$. The dof of $\chi^2_{[\ga,\al,\be]}$ is 3 = 14(observations for $\al$, $\be$ and $\ga$)-3(fitting parameters)-8(independent ISs on $\al$ and $\ga$).

\section{\label{2DKP of AB}2D King relation with $(\al,\be)$ transitions}

The combination $(\al,\be)$ is the same as that investigated in Ref.~\cite{Counts2020}.
Here we perform the analysis for this combination to check the consistency of our analysis with that in Ref.~\cite{Counts2020}.
The minimum $\ch^2$ and the corresponding $p$-value are
\begin{align}
  \ch_{[\al,\be]}^2 =15.37 \quad (p=4.0\times 10^{-3}) .
\end{align}
The best-fit parameters are $F_{\be\al}=1.01141006(86)$ and $K_{\be\al}=120.160(23)$~GHz$\vdot$u.
They are consistent with the results shown by Ref.~\cite{Counts2020}, $F_{\be\al}'=1.01141024(86)$ and $K_{\be\al}'=120.208(23)$~GHz$\vdot$u.
Note that the central values are slightly different from each other because we have used different values for the inverse mass differences. 
We then introduce the QFS or the PS in Eq.~\eqref{Eq2d_NL} as higher-order ISs.
Including an additional source of the ISs, one of four dof is consumed by the additional fit parameter $H_{\be\al}$. 
Their $\ch^2$ minima are
\begin{align}
  \ch^2_{[\al,\be]}(\text{QFS}) =& 4.3 \quad (p= 0.23), \\
	\ch^2_{[\al,\be]}(\text{PS})   =& 5.4 \quad (p= 0.15). \label{EqMitsm}
\end{align}
Following Ref.~\cite{Counts2020}, we minimize $\ch^2$ including both of the higher-order ISs,
\begin{align}
  \ch^2_{[\al,\be]}(\text{QFS}, \text{PS}) =& 4.0 \quad (p= 0.13),
\end{align}
where $p$-value is calculated with the dof of 2.
From the PS fit, we obtain the bound on the new physics coupling $3.8\times 10^{-11}\leq (-1)^{s+1} y_e y_n/(\hbar c) \leq 1.7\times 10^{-10}$ for the new particle mass of 10~eV at 95\% C.L.
The summary of the above fit results and the minimum $\ch^2$ are shown in Table~\ref{TabAB}.

\renewcommand{\arraystretch}{1.6}
\begin{table*}[htb]
\tabcolsep = 0.2cm
\centering
 \caption{
  Best-fit parameters with transition pairs $(\la_1,\la_2)= (\al, \be)$. The error of each fit parameter is evaluated as $1 \si$.}
\begin{tabular}{crrrrr}
\hline \hline
 Sources & \multicolumn{1}{c}{$\ch^2$ ($p$-value)} &  \multicolumn{1}{c}{$F_{\be\al}$} & \multicolumn{1}{c}{$K_{\be\al}$ (GHz$\vdot$u)}&\multicolumn{1}{c}{ $H_{\be\al, \text{QFS}}$  (kHz/fm$^4$)} &\multicolumn{1}{c}{ $ H_{\be\al, \text{PS}}$  (kHz)} \\
 \hline
           
	       QFS & 4.3 (0.23)              & 1.0114016(27)  & 120.381(70) & 71(21) & 0 \\
	        PS & 5.4 (0.15)              & 1.0114018(27)  & 121.55(44)  &      0 & 40(13) \\
	QFS and PS & 4.0 (0.13)              & 1.0114012(28)  & 120.77(81)  & 52(45) & 13(26) \\
\hline \hline
\end{tabular}

\label{TabAB}
\end{table*}

Here we consider the origin of the minimum values of  $\ch^2$ obtained in the above analysis.
The ISs should satisfy the transitive consistency condition $\nu^{A'A}_\la =\nu^{A'A_0}_\la-\nu^{A A_0}_\la$ for any transitions by definition.
With this condition, the shifts of some isotope pairs can be given by combinations of other pairs.
In the case of Ref.~\cite{Counts2020}, three isotope pairs $(170, 172)$, $(172, 174)$ and $(170, 174)$ are in this situation.
The minimum contributions from the condition to $\ch^2$ is given by
\begin{align}
  \ch^2_{C\la} (A_0;A_1,A_2) = \frac{(\nu^{A_1 A_0}_\la-\nu^{A_2 A_0}_\la -\nu^{A_1 A_2}_\la)^2}{(\si^{A_1 A_0}_\la)^2+(\si^{A_2 A_0}_\la)^2+(\si^{A_1 A_2}_\la)^2}.
\end{align}
Using the results of Ref.~\cite{Counts2020}, we find the lower limit of $\ch^2$ when we include the transitions $\al$ or $\be$,
\begin{align}
  \ch^2_{C\al}(172;170,174) =& 0.1262,\label{EqIc1} \\
  \ch^2_{C\be}(172;170,174) =& 3.911, \label{EqIc2} \\
  \ch^2_{C\al}(172;170,174) +\ch^2_{C\be}(172;170,174) =& 4.037.  \label{EqIc3}
\end{align}
These are the theoretical minima of $\chi^2$ in our analyses with $\alpha$ and/or
$\beta$ included.

\section{\label{finite value} Analysis of the nonlinearity of generalized King plot assuming QFS and PS}
\begin{figure}
\includegraphics[width=0.95\linewidth]{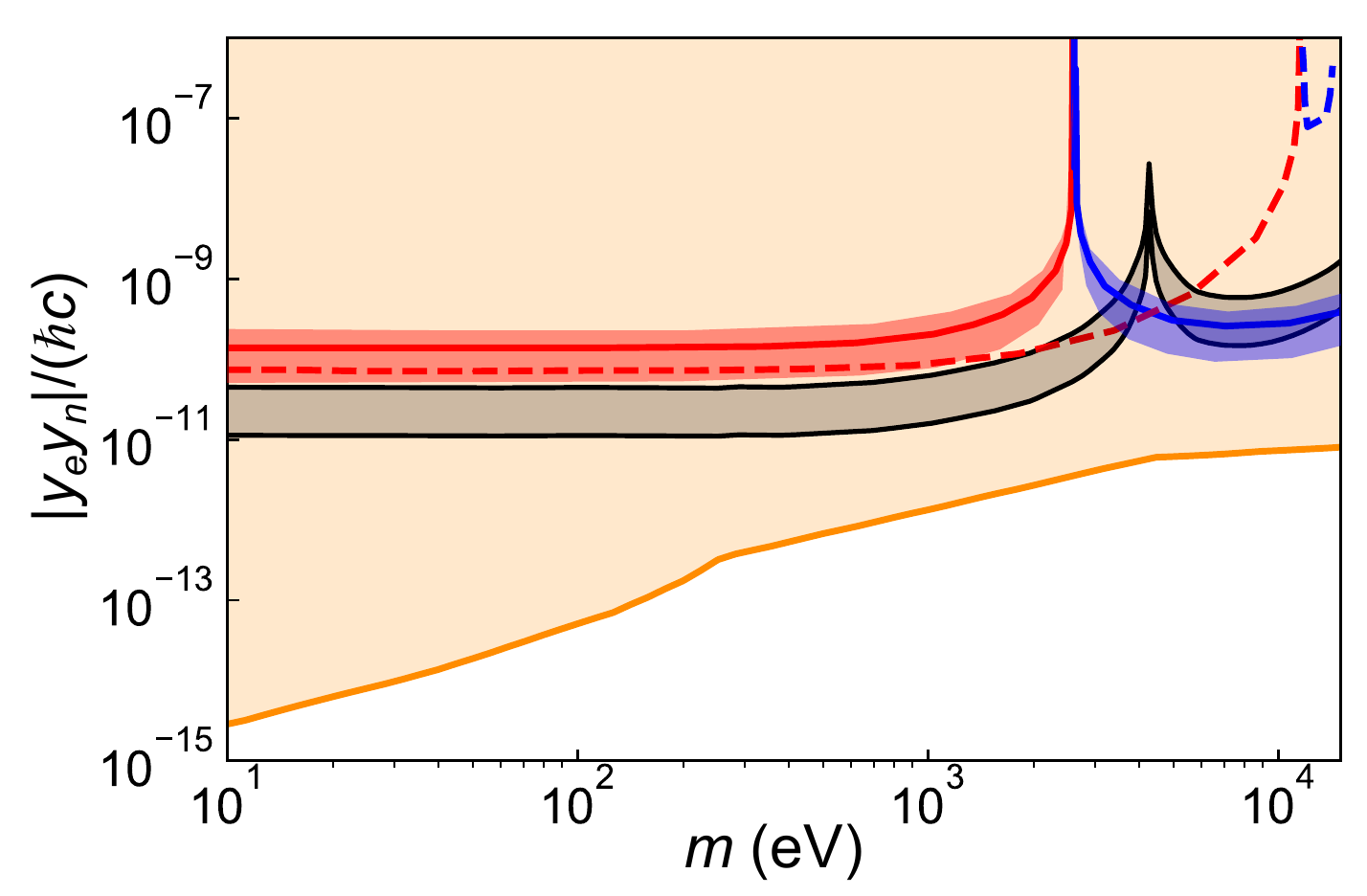}
\caption{Product of couplings $|y_ey_n|$ of a new boson as a function of the mass $m$. The black shaded region represents the 95\% confidence interval of the new physics coupling obtained from the fit using Eq.~\eqref{Eq3d_NL} with $h=h_\text{PS}$. For comparison, the constraint from electron anomalous magnetic moment $(g-2)$ measurements combined with neutron scattering measurements is shown as the orange shaded region, as in Fig.~\ref{NPBound}. In addition, the favored regions and constraints from Yb$^{+}$ analysis \cite{Counts2020} are shown as red and blue lines and shaded regions, in exactly the same manners as in Ref.~\cite{Counts2020}.}
\label{NPBound_finite}
\end{figure}

Here we suppose that the QFS and PS are the two distinct higher-order ISs involved in the data, as assumed in Refs.~\cite{Counts2020,Berengut2020}.
Furthermore, we consider the particular case where the QFS is eliminated in the 3D King plot construction and the PS remains as the origin of the nonlinearity in the 3D King relation.
This corresponds to the case of Eq.~\eqref{Eq3d_NL} with the PS-origin nonlinearity term given by $h_\text{PS}$.
From this analysis, we find that the $\chi^2$ minimum reaches the theoretical minimum, and we can determine the best-fit values for $h_\text{PS}$ as well as for the coefficients $f_{\al}$, $f_{\ga}$, and $k_{\mu}$ given in Table~\ref{TabABC}.
The $h_\text{PS}$ is expressed as $h_\text{PS} =\al_\text{NP}(X_{\be} -f_{\al}X_{\al} -f_{\ga} X_{\ga})$. See Appendix~\ref{Formulation of h} for the detail. 
We employed the electronic factor of the $\beta$-transition $X_{\beta}$ which is calculated by the configuration interaction method in Ref.~\cite{Counts2020}. The other factors $X_{\alpha, \gamma}$ are reconstructed so as to reproduce the results of Ref.~\cite{Berengut2020}. We evaluated $f_{\alpha, \gamma}$ also with other electronic factors given by Ref.~\cite{Berengut2020}. 
Thus, from the best-fit value of $h_\text{PS}$, we can determine the favored region of the coupling and mass of the new boson, as shown as the black shaded region in Fig.~\ref{NPBound_finite}.
For instance, the favored region is $1.1\times 10^{-11}\leq (-1)^{s} y_e y_n/(\hbar c) \leq 4.5\times 10^{-11}$ in the 95\% C.L. when the mass of the new boson is 10~eV. 
Note that the product of the couplings $y_e y_n$ is positive(negative) for $s=0$($s=1$) in the smaller mass side of the peak structure, namely $m\lesssim10$~keV. 
This peak structure is attributed to the cancellation of the electronic factors, and the sign of $y_e y_n$ changes across the peak. 
The suggested favored region, however, conflicts with the exclusion limit set by the other terrestrial experiment, obtained from the product of the individual bounds on the couplings with electron and neutron, as shown as the orange line in Fig.~\ref{NPBound_finite}.
Thus, we conclude that the QFS+PS assumption is not valid to describe the observed nonlinearities in the Yb/Yb$^+$ system.
\renewcommand{\arraystretch}{1.6}
\begin{table}[tb]
\tabcolsep = 0.1cm
\centering
 \caption{Best-fit parameters for 3D King relation with the PS term as an origin of the nonlinearity. 
The error of each fit parameter is evaluated as $1 \si$. 
}
\begin{tabular}{rrrrr}
\hline \hline
 \multicolumn{1}{c}{$f_\al$} &\multicolumn{1}{c}{$f_\ga$} & \multicolumn{1}{c}{$k_\mu$ (GHz$\vdot$u)}&\multicolumn{1}{c}{$h_\text{PS}$ (kHz)}\\
 \hline
 0.993(16) &  0.030(26) & 111.1(9.0) & 50(15)\\
\hline
\end{tabular}
\label{TabABC}
\end{table}

\section{\label{Formulation of h}Explicit forms of $h_\text{PS}$ and $h_\text{QFS}$}

Here we consider two distinct higher-order sources, defined as $I_\la\de\eta^{A'A}$ and $J_\la\de\zeta^{A'A}$, and the IS is expressed as
\begin{align}
   \nu_{\la}^{A'A}=K_{\la} \de\mu^{A'A} +F_{\la} \de\avg{r^2}^{A'A} 
   +I_\la\de\eta^{A'A} +J_\la\de\zeta^{A'A}.
\end{align} 
When we eliminate $\de\avg{r^2}^{A'A}$ and $\de\eta^{A'A}$ by combining the three transitions $(\la_1,\la_2,\la_3)$, the coefficients of the 3D King relation Eq.~\eqref{Eq3d_NL} are expressed as
\begin{align}
f_{\la_1} = \frac{F_{\la_2}I_{\la_3}-F_{\la_3}I_{\la_2}}{F_{\la_2}I_{\la_1}-F_{\la_1}I_{\la_2}},
\label{f1}
\end{align}
\begin{align}
f_{\la_2} = \frac{F_{\la_3}I_{\la_1}-F_{\la_1}I_{\la_3}}{F_{\la_2}I_{\la_1}-F_{\la_1}I_{\la_2}},
\label{f2}
\end{align}
\begin{align}
k_\mu = K_{\la_3}-f_{\la_1}K_{\la_1}-f_{\la_2}K_{\la_2},
\end{align}
\begin{align}
h = J_{\la_3}-f_{\la_1}J_{\la_1}-f_{\la_2}J_{\la_2}.
\end{align}
If $J_\la\de\zeta^{A'A}$ corresponds to the QFS or PS, the nonlinear terms $h_\text{QFS}$ and $h_\text{PS}$ are expressed as
\begin{align}
h_\text{QFS} = G^{(2)}_{\la_3}-f_{\la_1}G^{(2)}_{\la_1}-f_{\la_2}G^{(2)}_{\la_2},
\end{align}
\begin{align}
h_\text{PS} = \al_\text{NP}(X_{\la_3}-f_{\la_1}X_{\la_1}-f_{\la_2}X_{\la_2}).
\end{align}
Note that the coefficients in $h_\text{PS}$ and  $h_\text{QFS}$, shown in Eqs.~\eqref{f1} and \eqref{f2}, depend on the electronic factors of the eliminated terms $I_{\la_1}$, $I_{\la_2}$, and $I_{\la_3}$.

\section{\label{Consistency check} Possible consistency check in the nonlinearity fit}

In Appendix~\ref{finite value}, following Refs.~\cite{Counts2020,Berengut2020}, we describe the 3D King plot analysis under the assumption that the QFS and PS are two distinct higher-order ISs involved in the data.
This picture is not plausible to explain the observed deviations from the linearities because the given favored region is excluded by the other experiments.
In general, even if the data set is fit by some higher-order ISs well, they do not have to be the origins of the observed nonlinearity.
Here, we discuss a method to test the origins of the higher-order ISs.

Different from the argument in Appendix~\ref{finite value} where we attribute the PS to the source of the nonlinearity and the QFS is eliminated by the construction of 3D King plot, we here attribute the QFS, instead of the PS, to the source to explain the leftover nonlinearity and the PS is eliminated.
These two constructions should be treated on an equal footing, in principle.
This fit gives us the $\ch^2$ of the theoretical minimum, see Table~\ref{NewMethod} for the fit result.
In this case, the given fit coefficients can be calculated with the electronic factors $F_\la$, $K_\la$ and $X_\la$ $(\la \in \{\al, \be, \ga \})$ using the formulae shown in Appendix~\ref{Formulation of h}.
The coefficients calculated by the electronic factors given in Ref.~\cite{Counts2020} do not match the fit result for $m<1$~keV \cite{Note2}.
Table~\ref{NewMethod} shows the QFS fit result and the theoretical coefficients at $m=10$~eV. 
This means, as long as we use the numerical results given by Ref.~\cite{Berengut2020}, the original assumption to include only the QFS and the PS ($m<1$~keV) as the higher-order ISs is inconsistent with the data.
In the generalized King relation, this method helps us to test the consistency of some higher-order ISs with experimental results independent of other experimental bounds.


\renewcommand{\arraystretch}{1.6}
\begin{table}[tb]
\tabcolsep = 0.1cm
\centering
 \caption{
  Best-fit parameters for the 3D King relation with the QFS term as the origin of the nonlinearity. 
In the QFS fit, the error of each fit parameter is evaluated as $1 \si$. 
The second line is the coefficients at $m=10$~eV calculated by the electronic factors given by Ref.~\cite{Berengut2020}.
}
\begin{tabular}{crrrrr}
\hline \hline
  &  \multicolumn{1}{c}{$f_\al$} &\multicolumn{1}{c}{$f_\ga$} & \multicolumn{1}{c}{$k_\mu$ (GHz$\vdot$u)}&\multicolumn{1}{c}{$h_\text{QFS}$ (kHz/fm$^4$)}\\
 \hline
  QFS fit& 1.018(13) & -0.010(21) & 124.1(7.7) & 70(21)  \\
\hline
  \cite{Berengut2020} & 1.063 & -0.069 & 240.1 & -57 \\
\hline
\end{tabular}
\label{NewMethod}
\end{table}

\hspace{1cm}
\begin{acknowledgments}
We thank Ayaki Sunaga for fruitful discussions. 
We also thank Kantaro Honda for his contribution on stabilizing the clock laser frequency.
KO acknowledges support from the JSPS (KAKENHI grant number JP19J11413). 
This work of MT is supported in part by JSPS KAKENHI Grant No.~JP18K03621.
The work of YY is supported in part by the National Science Centre (Poland) the research Grant No.~2017/26/D/ST2/00490, and the Ministry of Science and Technology (MOST) of Taiwan and the
National Center for Theoretical Sciences (NCTS).
The experimental work was supported by the Grant-in-Aid for Scientific Research of JSPS(Nos.~JP17H06138, JP18H05405, JP18H05228, and JP21H01014), the Impulsing Paradigm Change through Disruptive Technologies (ImPACT) program, JST CREST (No.~JP-MJCR1673), and MEXT Quantum Leap Flagship Program (MEXT Q-LEAP) Grant No.~JPMXS0118069021.
\end{acknowledgments}



 \end{document}